\documentclass[manuscript]{aastex}

\usepackage{graphics,color}
\usepackage{epstopdf}
\definecolor{red}{rgb}{1,0,0}

\shorttitle{Asteroid Rotation Periods From iPTF}
\shortauthors{Chang et al.}
\begin{document}
\title{Asteroid Discovery and Light Curve Extraction Using the Hough Transform -- A Rotation Period Study for Sub-Kilometer Main-Belt Asteroids}

\author{Kai-Jie Lo\altaffilmark{1}, Chan-Kao Chang\altaffilmark{2}, Hsing-Wen Lin\altaffilmark{3}, Meng-Feng Tsai\altaffilmark{1}, Wing-Huen Ip\altaffilmark{2,4}, Wen-Ping Chen\altaffilmark{2}, Ting-Shuo Yeh\altaffilmark{2}, K. C. Chambers\altaffilmark{4}, E. A. Magnier\altaffilmark{4}, M. E. Huber\altaffilmark{4}, R. J. Wainscoat\altaffilmark{4}}

\altaffiltext{1}{Department of Computer Science and Information Engineering, National Central University, Jhongli, Taiwan}
\altaffiltext{2}{Institute of Astronomy, National Central University, Jhongli, Taiwan}
\altaffiltext{3}{Department of Physics, University of Michigan, Ann Arbor, Michigan 48109, USA}
\altaffiltext{4}{Space Science Institute, Macau University of Science and Technology, Macau}
\altaffiltext{5}{Institute for Astronomy, University of Hawaii, Honolulu, Hawaii 96822, USA}

\email{rex@astro.ncu.edu.tw}


\begin{abstract}
The intra-night trajectories of asteroids can be approximated by straight lines, and so are their intra-night detections. Therefore, the Hough transform, a line detecting algorithm, can be used to connect the line-up detections to find asteroids. We applied this algorithm to a high-cadence Pan-STARRS 1 (PS1) observation, which was originally designed to collect asteroid light curves for rotation period measurements \citep{Chang2019}. The algorithm recovered most of the known asteroids in the observing fields and, moreover, discovered 3574 new asteroids with magnitude mainly of $21.5 < w_{p1} < 22.5$ mag. This magnitude range is equivalent to sub-kilometer main-belt asteroids (MBAs), which usually lack of rotation period measurements due to their faintness. Using the light curves of the 3574 new asteroids, we obtained 122 reliable rotation periods, of which 13 are super-fast rotators (SRFs; i.e., rotation period of $< 2$ hr). The required cohesion to survive these SFRs range from tens to thousands of $Pa$, a value consistent with the known SFRs and the regolith on the Moon and Mars. The higher chance of discovering SFRs here suggests that sub-kilometer MBAs probably harbor more SFRs.
\end{abstract}


\section{Introduction}
Finding asteroids is the fundamental step for asteroid research. With the advances in wide-field cameras, robotic observations, and computing resources, several telescopes have been dedicated to asteroid discovery, such as the Catalina Sky Survey (CSS)\footnote{http://www.lpl.arizona.edu/css/}, the Pan-STARRS 1 \citep{Chambers2016}, the Asteroid Terrestrial-impact Last Alert System \citep[ATLAS;][]{Tonry2018}, and the Chinese Near-Earth-Object Survey Telescope (CNEOST)\footnote{http://www.cneost.org/}. To optimize the discovery rate of asteroid, each project customizes its survey rates (i.e., scanning sky area over a period of observation time) to a certain cadence that meets the basic requirement of discovering an asteroid, such as three detections within a night or several detections over a few nights for a single asteroid discovery. Therefore, each project develops its own asteroid-detecting pipeline and optimizes it to the scientific interests and survey strategy. In some cases, the pipeline does not utilize all the detections for some reasons. The most common case is a filter to exclude the low signal-to-noise ratio detections, which usually contain many noises. Nevertheless, if a survey was conducted using a relatively short cadence, asteroids could be much easier to be distinguished from the background noises and field stars. A good example of high-cadence observation is the rotation period survey for MBAs carried out in October, 2016 using the PS1 \citep{Chang2019}, in which eight PS1 fields were continuously scanned using $\sim 10$ min cadence over six straight nights. From the survey, the intra-night detections of asteroids appeared as straight lines due to their approximately linear motion during the short period of time, which obviously stand out from the stationary sources and noises. Therefore, we can make use of the Hough Transform to connect these intra-night line-up detections to find asteroids. The Hough Transform was originally proposed by \citet{Hough1959} and then further utilized by \citet[][i.e., generalized Hough transform]{Duda1972}. This method has been widely used to extract the linear features on two-dimension images and recently started to be applied to astronomical images to detect cosmic rays \citep{Keys2010}, streaks produced from fast-moving small solar system bodies (SSSBs) \citep[][]{Virtanen2014, Kim2016, Bektesevic2017, Nir2018}, and long tracks of space debris \citep{Hickson2018}. Instead of working on images, we applied the Hough Transform to the source catalogs obtained from the aforementioned intra-night observations. The implementation of the Hough Transform is relatively simple and, moreover, it does not have to exclude low signal-to-noise detections in advance. Therefore, we were able to utilize the data set much closer to its actual limiting magnitude (i.e., $\sim 22.5$ mag) and found more than three thousand asteroids mainly of $21.5 < w_{p1} < 22.5$ mag. With this magnitude, a study of rotation period for sub-kilometer MBAs was carried out for the first time\footnote{Most sub-kilometer sized or smaller objects with available rotation periods are all near-earth asteroids.}.


In this work, the PS1 observation and the data set are briefly described in Section 2, the Hough Transform is illustrated in Section 3, the result of asteroid recovery, discovery, and rotation period analysis are presented in Section 4, and a short summary is given in the end.

\section{The PS1 Observation and the Data Set}\label{dataset}
The PS1 was designed to discover small solar system bodies and, especially, those potentially hazardous objects. It is a 1.8 m Ritchey-Chretien reflector located on Haleakala, Maui equipped with the Gigapixel Camera having a field of view of 7 deg$^2$. Six filters are used, including $g_{P1}$ ($\sim 400-550$ nm), $r_{P1}$ ($\sim550-700$ nm), $i_{P1}$ ($\sim 690–820$ nm), $z_{P1}$ ($\sim820-920$ nm), $y_{P1}$ ($>920$ nm), and $w_{P1}$ (i.e., combination of $g_{P1}$, $r_{P1}$, and $i_{P1}$), in which the $w_{P1}$ filter is especially designed for moving object discovery \citep{Kaiser2010, Tonry2012, Chambers2016}. The images obtained by the PS1 are processed by the Image Processing Pipeline \citep[IPP;][]{Chambers2016, Magnier2016a, Magnier2016b, Magnier2016c, Waters2016} to provide source catalogs as the final product.

During October 26-31, 2016, a high-cadence observation was carried out using the PS1 to collect a large sample of asteroid light curves for rotation period measurement \citep{Chang2019}. The observation was conducted using a cadence of $\sim 10$ minutes to repeatedly scan eight consecutive PS1 fields in $w_{P1}$ band. Therefore, each field had $\sim 30$ exposures each night, except for the last two nights which only had few exposures were taken due to unstable weather. Moreover, several exposures in each $g_{P1}$, $r_{P1}$, $i_{P1}$, and $z_{P1}$ band were also taken in the first night between the exposures of $w_{P1}$ band to obtain asteroid colors. Table~\ref{obs_log} shows the summary of the observation. Although $\sim 3500$ objects, mostly $< 21.5$ mag, have been discovered and reported to the Minor Planet Center\footnote{https://www.minorplanetcenter.net/} using this observation, a lot of unknown objects close to the actual limiting magnitude of $\sim 22.5$ have not been found yet. This high-cadence observation is suitable for asteroid discovery using the Hough Transform. Therefore, the source catalogs obtained from the observation were used as our running cases.

\section{The Algorithm}
\subsection{The Hough Transform}
\begin{figure}
\epsscale{1}
\plotone{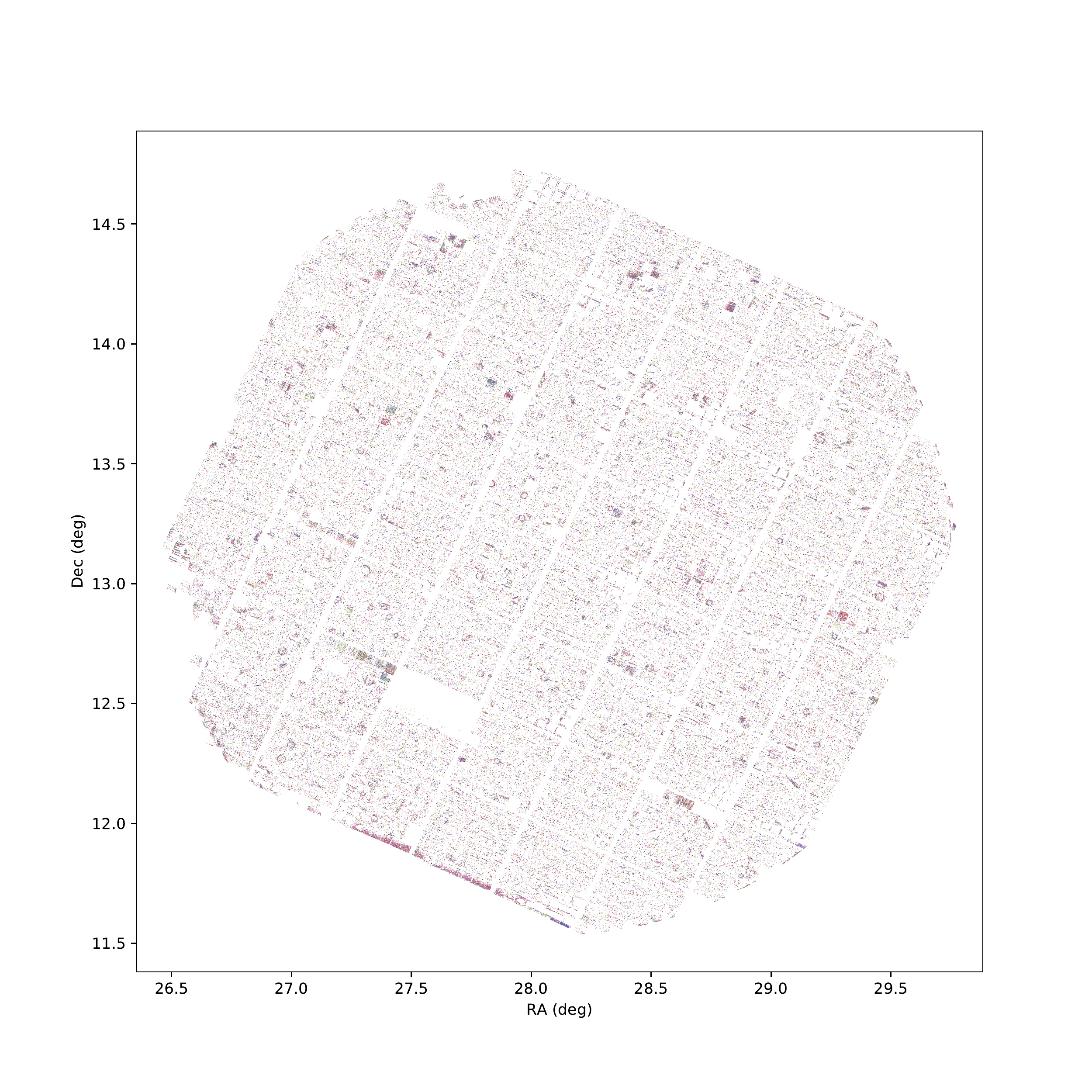}
\caption{The master frame. The detections, taken from the source catalogs of the observations for the field 1 carried out on October 26, 2016, are stacked into a single frame in which the observing epochs are color coded.}
\label{masterframe}
\end{figure}

\begin{figure}
\epsscale{1}
\plotone{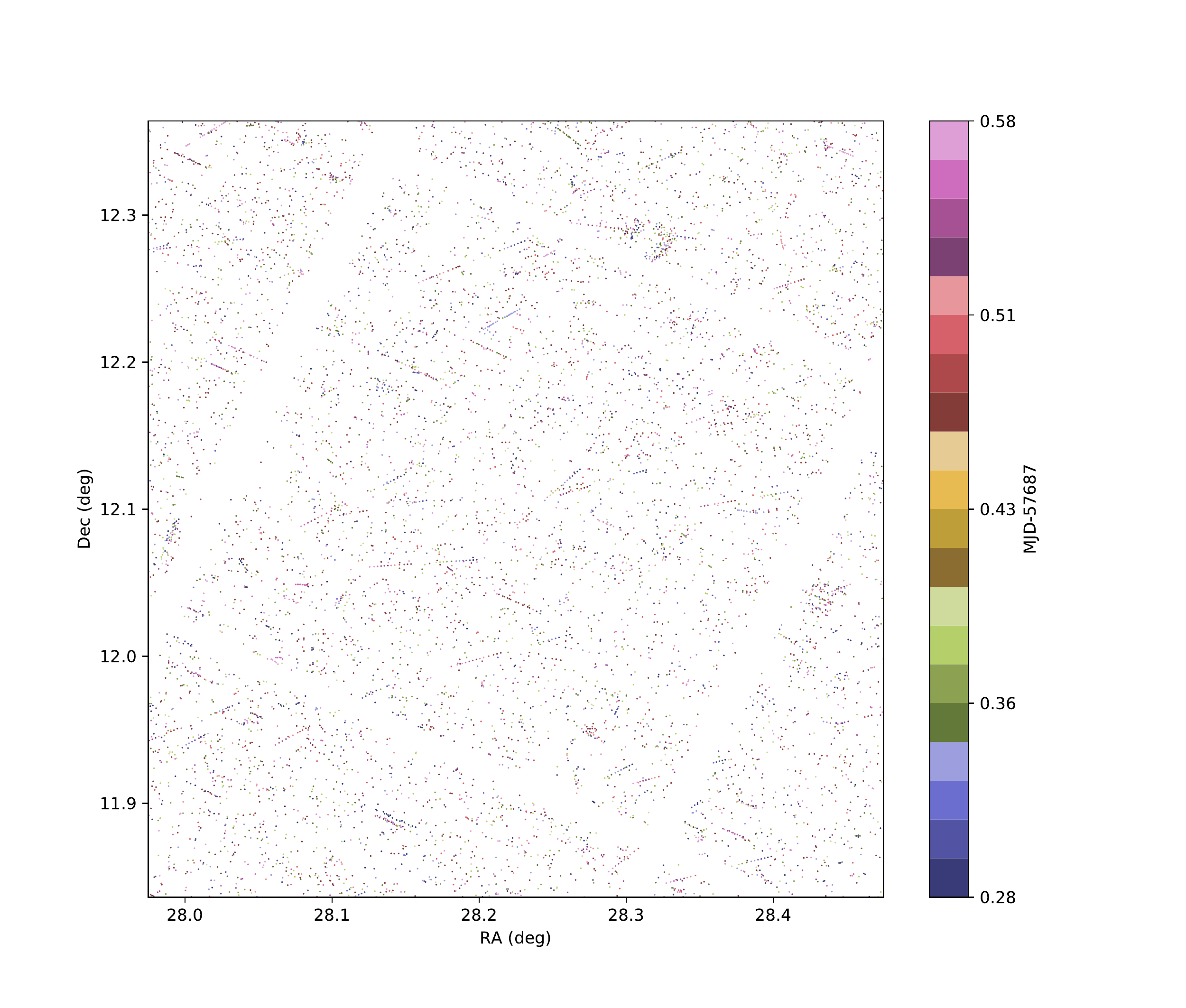}
\caption{A zoom-in view of the Figure~\ref{masterframe}, where the detections of asteroids shown as straight lines with correct time sequence.}
\label{masterframe01}
\end{figure}

The intra-night trajectory of an asteroid can be approximated by a straight line, and so do its detections obtained from intra-night observations. Therefore, we can find asteroids through these line-up intra-night detections. The situation is illustrated in Figure~\ref{masterframe} and~\ref{masterframe01}, in which all the detections, obtained from all the source catalogs of the intra-night observations for one particular field, are stacked into a single frame (hereafter, the master frame). On the master frame, we see that the intra-night detections of asteroids on that field appear as straight lines with correct time sequence.

\begin{figure}
\plotone{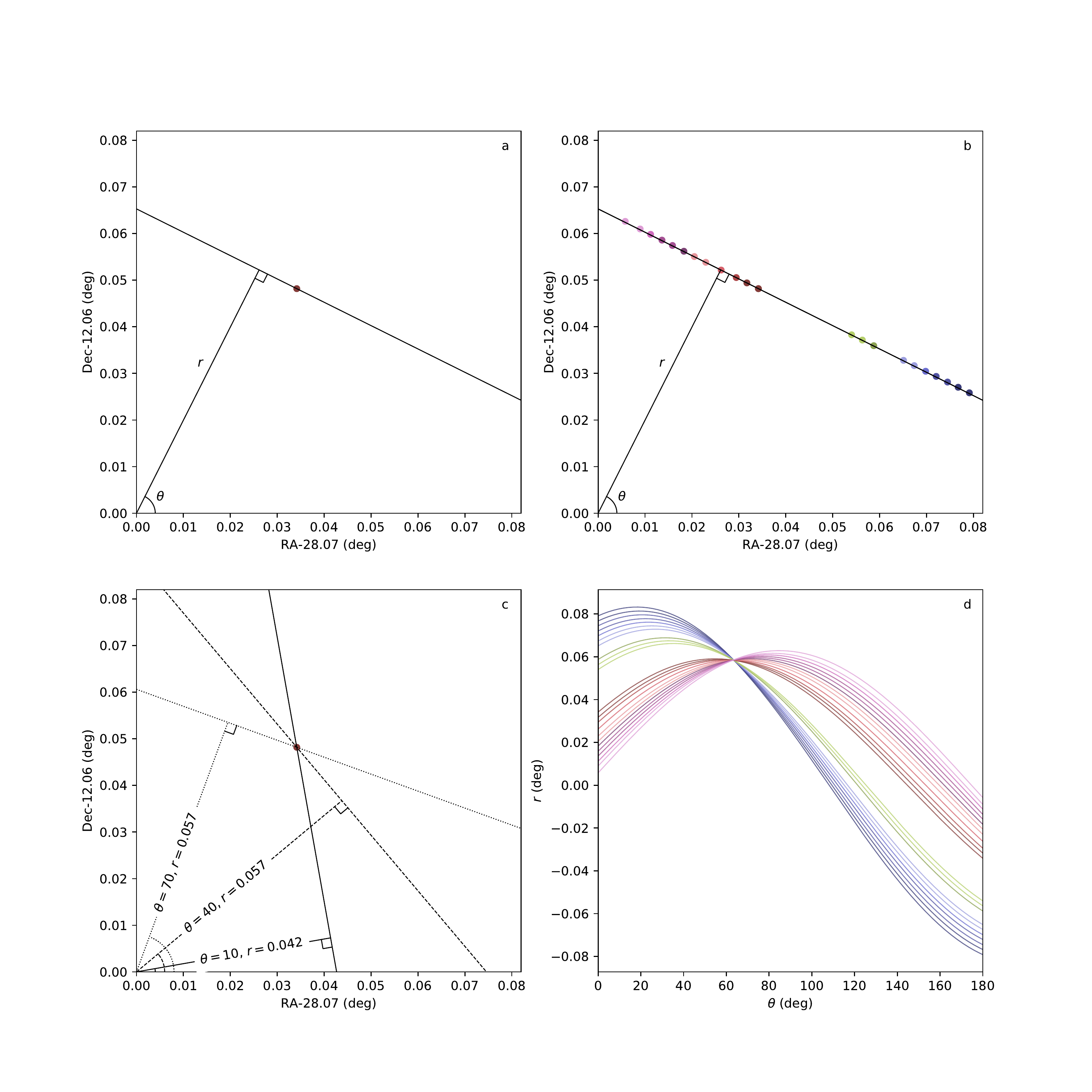}
\caption{The illustration of asteroid discovery using the Hough Transform. Plot a: a particular line, which passes through a detection, is expressed by ($\theta$, $r$) on the master frame. Plot b: if the detections on the master frame belonging to the same line, they would share common ($\theta$, $r$). Plot c: The parameter space of ($\theta$, $r$) is explored for each detection on the master frame. Plot d: the ($\theta$, $r$) curves of the detections on the plot b, in which the curves and the corresponding detections use the same colors. Since the curves converge to ($\theta$, $r$) = (63.4, 0.057), it suggests that these detections belong to that particular line.}
\label{hough}
\end{figure}

Using the Hough Transform, a line on a two dimension image can be expressed in the Hesse normal form as
\begin{equation}
  r = x \cos \theta + y \sin \theta,
\end{equation}\label{hg_eq}
where ($x$, $y$) is the coordinate of the pixel on the image, $r$ is the distance to the closet point of the line to the reference point, and $\theta$ is the angle between this line and the $x$ axis (see Figure~\ref{hough}a). In our case, the ($x$, $y$) is the sky coordinate, RA and Dec. When a group of detections belong to the same straight line, they would share common/similar ($\theta$, $r$) (see Figure~\ref{hough}b). Therefore, we can explore the parameter space of ($\theta$, $r$) for all the intra-night detections on the master frame (see Figure~\ref{hough}c) and search for those sharing common ($\theta$, $r$) to locate MBAs (see Figure~\ref{hough}d.

\subsection{The Implementation}\label{HTimp}
\subsubsection{Detection clean-up and master frame generation}\label{cleanup}
Before stacking the intra-night detections of a field to create the mater frame, any detection was first removed from a source catalog if it is on the noisy area of the PS1 detector chip \citep[][see Figure 4 therein]{Waters2016} or around any stationary source within 2\arcsec. Then, the remaining intra-night detections were stacked into a master frame. According to the bulk motion of MBAs within 10 minutes, one asteroid detection was expected to have at least another companion on the master frame. Therefore, any detection on the master frame was further removed if it does not have any other neighbor within 1.5 to 30\arcsec. Typically, each field each night has $\sim 260000$ detections left on the master frame.

\subsubsection{Hough Transform calculation and segment identification on the master frame}\label{sec_segment}
After the clean-up step, the Hough Transform, Equation~(\ref{hg_eq}), was applied to each detection on the master frame, in which $r$ was calculated by exploring $\theta$ in a range of 1 to 179 degree with a step of 0.1 degree. Next, we searched for the detections sharing common ($\theta$, $r$) to identify the line-up detections (hereafter, segment). Here, a common ($\theta$, $r$) means the same value of $\theta$ and the same value of $r$ till the fourth digit after the decimal point (i.e., within 0.0001 degree difference)\footnote{The 0.0001 degree or 0.36\arcsec~difference in $r$ is considered as the uncertainty of a detection coordinates, which is much smaller than the average seeing, $\sim 1.2$\arcsec, during the observations.}. Only the segments with five or more detections were passed to the following processes. Typically, it takes $\sim 25$ min to complete this step for a master frame on a computer cluster of 5 nodes\footnote{The computer cluster contains 5 identical nodes, in which one is for job distribution and the other four are for computation. Each node has a 4-core Intel i7-2600@3.40 GHz cup and 16 GB memory.}.

\subsubsection{Segment combination on the master frame}
In some cases, the neighboring segments on the master frame share one or more identical detections. When the difference in $\theta$ between these segments is less than 0.1 degree, they are combined into a new segment, otherwise only the longest segment was remained. Moreover, two segments on the master frame are combined if they have the following situation: (a) their mean locations can be mutually predicted within 7\arcsec, (b) their difference in $\theta$ is less than 0.1 degree, and (c) their time spans have no overlap. A detection is further excluded from a segment when it has incorrect epoch in the time sequence (i.e., simple increase/decrease in observation epochs) or has a magnitude $3 \sigma$ away from the mean magnitude.

\subsubsection{Segment Linkage across the master frames}
\begin{figure}
\plotone{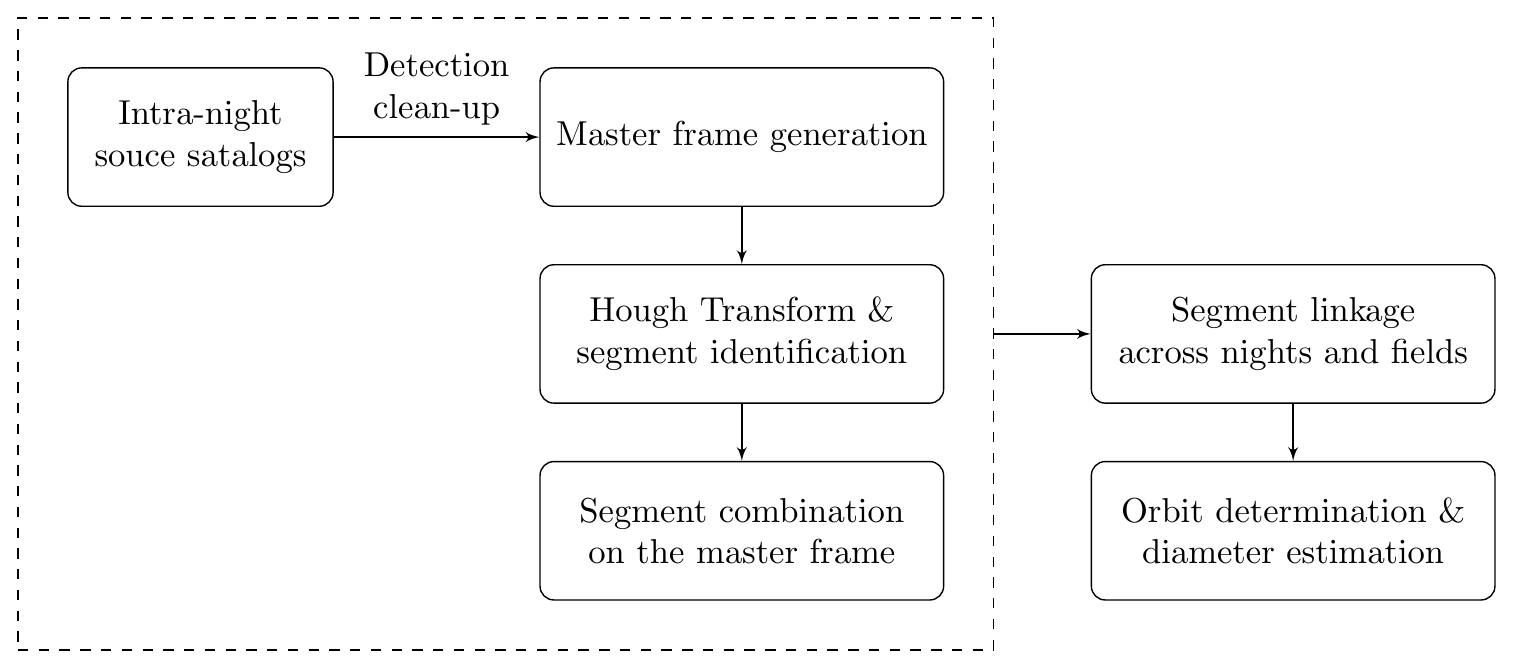}
\caption{The flow chart of the Hough Transform implementation.}
\label{flowchart}
\end{figure}

To connect segments across different nights and fields, a linkage was performed to all the master frames. If two segments can mutually predict each other's mean locations within 60\arcsec~and have a difference in $\theta$ less than 0.1 degree, they are identified as the same asteroid. In the end, the light curves of the remaining segments were extracted for further analysis. The flow chart of the Hoguh Transform implementations is given in Figure~\ref{flowchart}

\subsection{Orbital Determination and Diameter Estimation}
\begin{figure}
\plotone{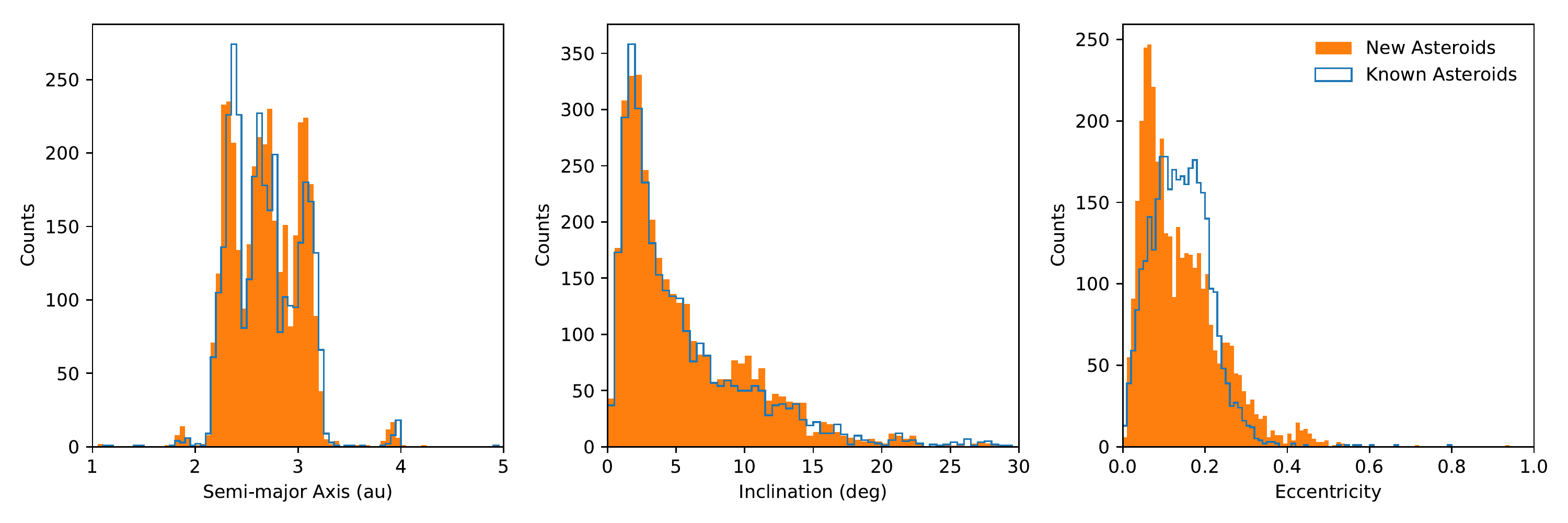}
\caption{The orbital distributions of the new asteroids and the known ones in the observing field.}
\label{orbit}
\end{figure}

To determine the preliminary orbits and absolute magnitudes in $w_{P1}$ band of the newly discovered asteroids, the {\it{Find\_Orb}} was adopted\footnote{https://www.projectpluto.com/fo.htm. Note that the absolute magnitude was calculated simply assuming a slope of 0.15 in the $H-G$ system \citep{Bowell1989}.}. Figure~\ref{orbit} shows the orbital distributions of the new asteroids along with the known ones in the observing fields. We see that the distributions are similar in the semi-major axis and inclination and, however, the new asteroids has a mean eccentricity slightly smaller than the known. This is because we only have four-straight-night arcs at the opposition, which provides relatively better constraints on the semi-major axis and inclination\footnote{The medium uncertainties in our semi-major axis, inclination, and eccentricity are $\sim 7\%$, $\sim 45\%$, and $\sim 84\%$, respectively.}.

To estimate the diameters of the new asteroids, we used
\begin{equation}\label{dia_eq}
  D = {1329 \over \sqrt{p_{v}}} 10^{-H_{w}/5},
\end{equation}
where $H_w$ is absolute magnitude in $w_{p1}$ band, $D$ is diameter in kilometer, $p_v$ is geometric albedo in $V$ band, and 1329 is the conversion constant. We assumed $p_v = 0.20$, 0.08, and 0.04 for asteroids in the inner ($2.1 < a < 2.5 AU$), mid ($2.5 < a < 2.8$ AU) and outer ($a > 2.8$ AU) main belts, respectively \citep{Tedesco2005}\footnote{The conversion from $H_w$ to $H_v$ requires color information, which is not available for our samples. Therefore, $H_w$ was used in Equation~(\ref{dia_eq}) instead of $H_v$. Typically, $H_v$ is $\sim 0.1$ to 0.3 mag fainter than $H_w$ for MBAs when applying the transformation listed in \citet{Tonry2012} and assuming a $g-r$ color varying between 0.3 - 0.8 \citep{Ivezic2001}. This magnitude difference makes $\sim 10\%$ difference in diameter estimation. In addition, the median uncertainty of our semi-major axes is about $\sim 7\%$. This makes a difference in $H_w \sim 0.4$ mag, corresponding to an additional uncertainty in diameter $\sim 20\%$. Considering this two factors, a typically uncertainty in our diameter estimation could be $\sim 30$ to $40\%$. However, this uncertainty does not take the assuming albedos into account, which could make several times difference in diameter estimation.}.

\section{Asteroid Recovery and Discovery}\label{sec_ast_rev}
To check the recovery rate of the Hough Transform, we used the known asteroids as of January 6, 2018 to compare with our result. The ephemerides of the known asteroids in the observing fields were obtained from the {\it JPL/HORIZONS} system, and then a cross match was performed against the PS1 source catalogs to extract their detections using a radius of 2\arcsec (hereafter, ephemeris-matching method). If an asteroid have five or more detections, we defined it as a successful recovery. In our observing fields, there were 3870 known asteroids recovered by the ephemeris-matching method, out of which 3819 were also recovered by the Hough Transform. For the asteroids lost by the Hough Transform, we found that the detections of these objects remaind on the master frames are relatively small and, therefore, they have a lower chance to be recovered by the Hough Transform. Two reasons account for this: (a) the detection number is intrinsically small (i.e., $\sim 80\%$ of them have $< 20$ detections using the ephemeris-matching method), and (b) the majority of the detections have been removed from the master frames in the clean-up step (i.e., $\sim 20\%$ of them have tens of detections matched using the ephemeris-matching method).

\begin{figure}
\plotone{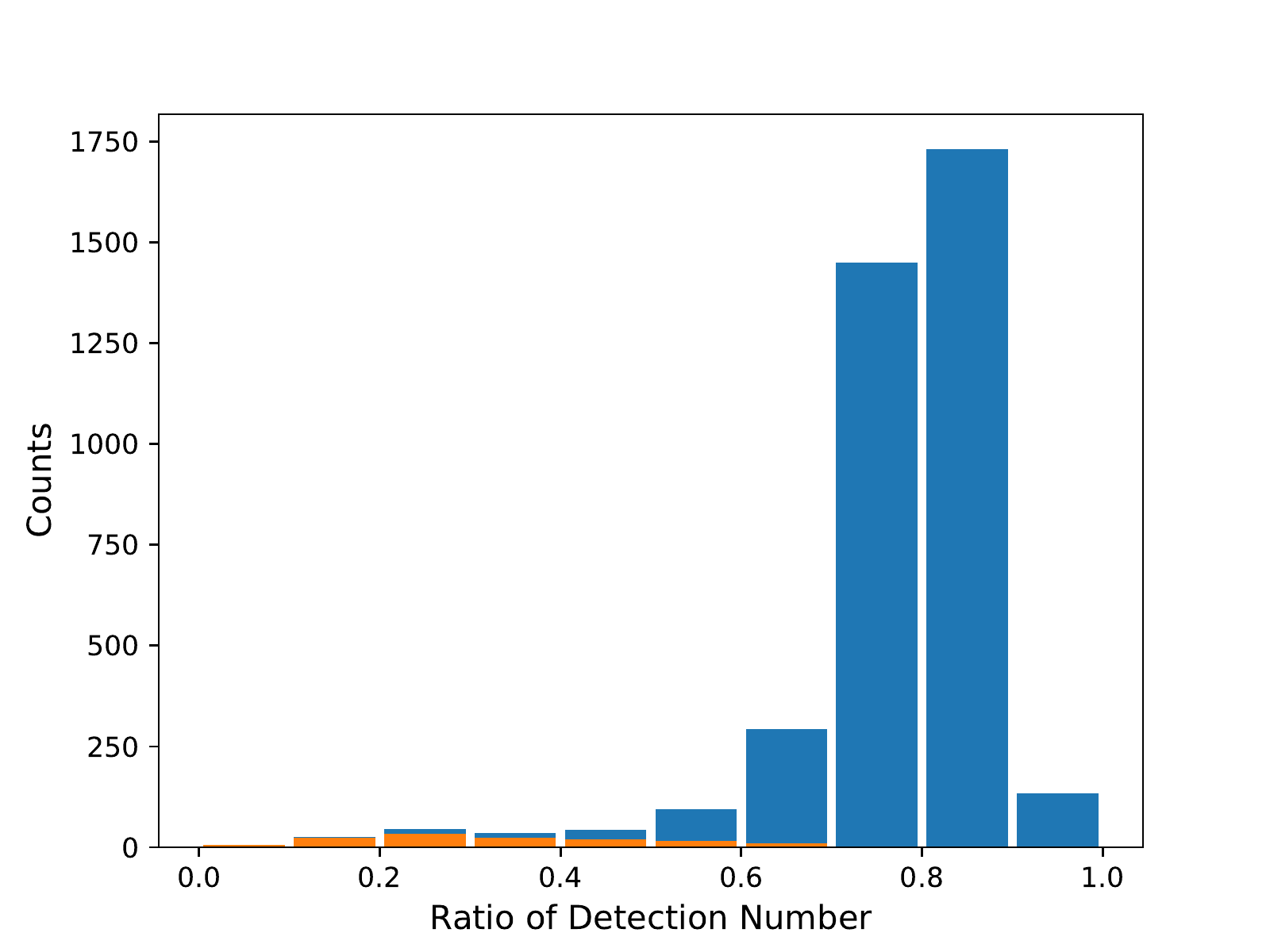}
\caption{The distribution of the ratio of the detection number recovered by the Hough Transform to that by the ephemeris-matching method, where the orange means that the known asteroids have two or more Hough Transform segments which are not probably connected in the linkage steps.}
\label{det_ratio}
\end{figure}

We then compare the detection number for the known asteroids recovered by the both methods. Figure~\ref{det_ratio} shows the distribution of the ratio of the detection number recovered by the Hough Transform to that extracted by the ephemeris-matching method, where we see that the Hough Transform has smaller detection number in average. This is because (a) a relatively strict criterion has been applied in the definition of a segment in the Hough Transfrom (i.e., 0.0001 degree difference in $r$ for a segment which is much smaller than the 2\arcsec~radius used in ephemeris-matching method; see Section~\ref{sec_segment}), and (b) some detections have been removed from the master frame in the clean-up step (see Section~\ref{cleanup}). However, a small portion of the known asteroids have much smaller ratio of detection number (i.e., $< 50\%$). We found that these low-ratio asteroids have multiple segments in the Hough Transform, which were not probably connected in the linking steps to become long segments (see the orange color in Figure~\ref{det_ratio}).

\subsection{Asteroid Discovery}
\begin{figure}
\plotone{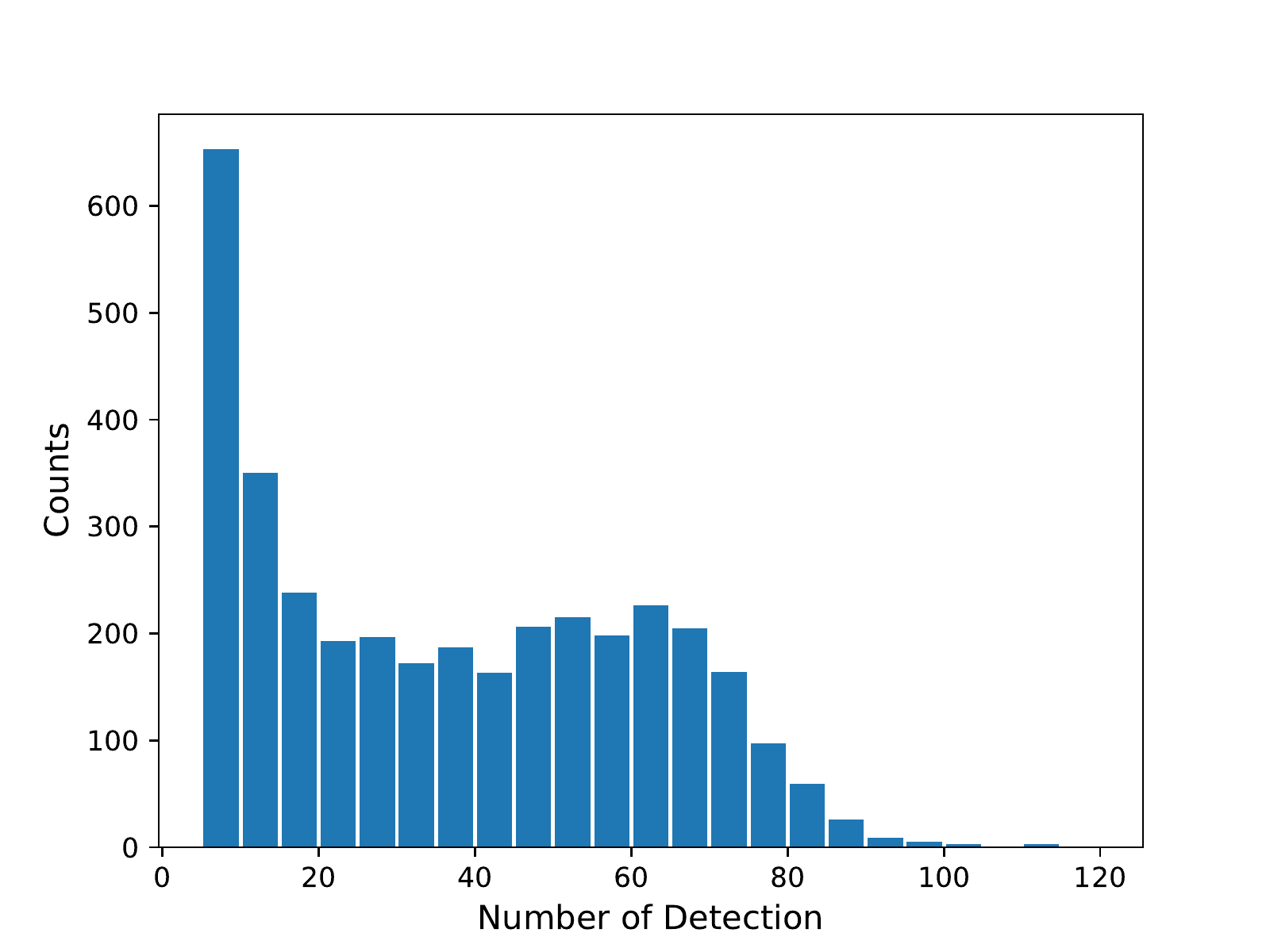}
\caption{The distribution of the detection number for the new asteroids.}
\label{det_number}
\end{figure}

\begin{figure}
\plotone{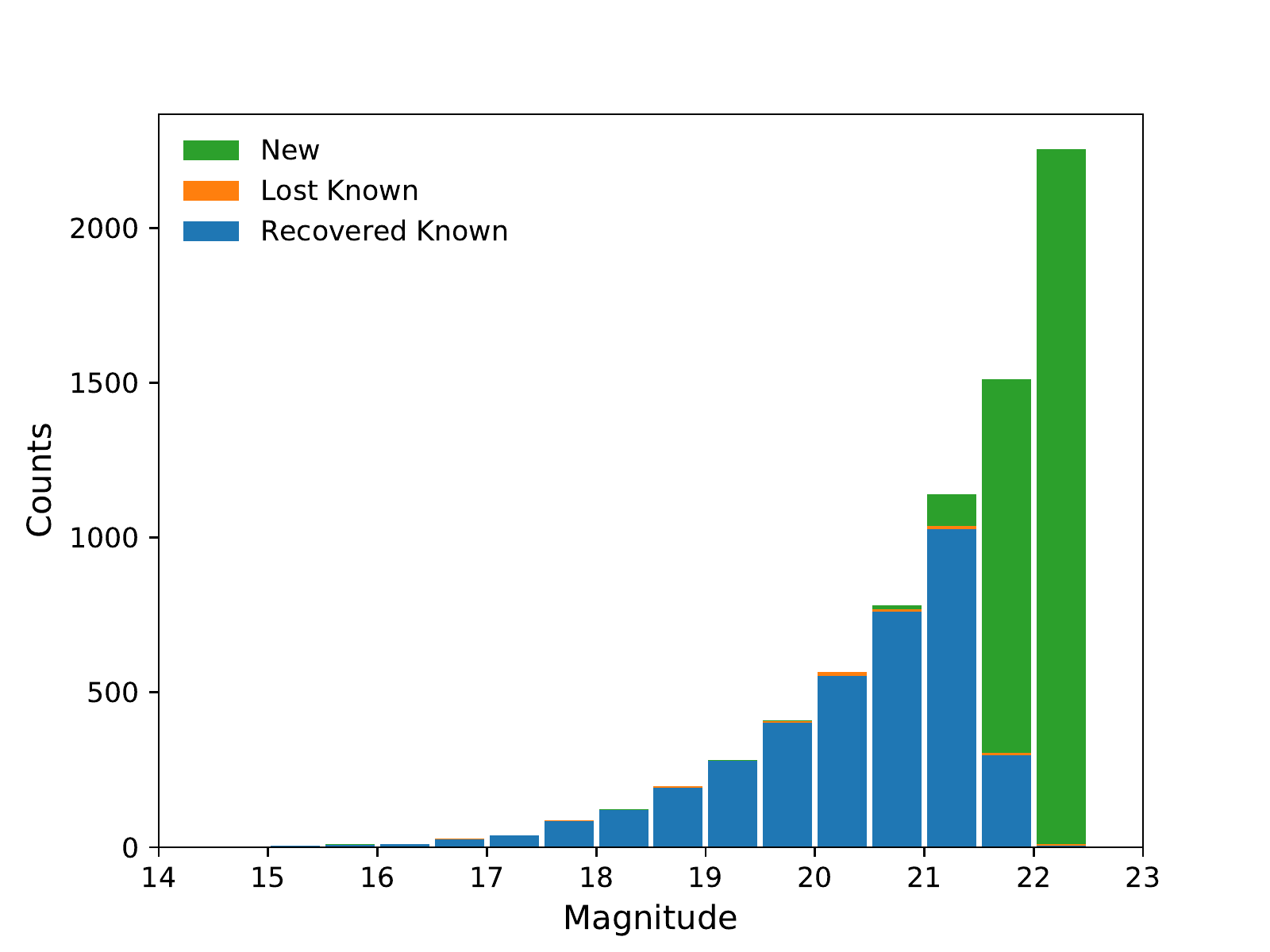}
\caption{The magnitude distribution of the asteroids in the observing fields. The blue and orange are the known asteroids recovered and lost by the Hough Transform, respectively, and the green is the new asteroids.}
\label{maghist}
\end{figure}

\begin{figure}
\plotone{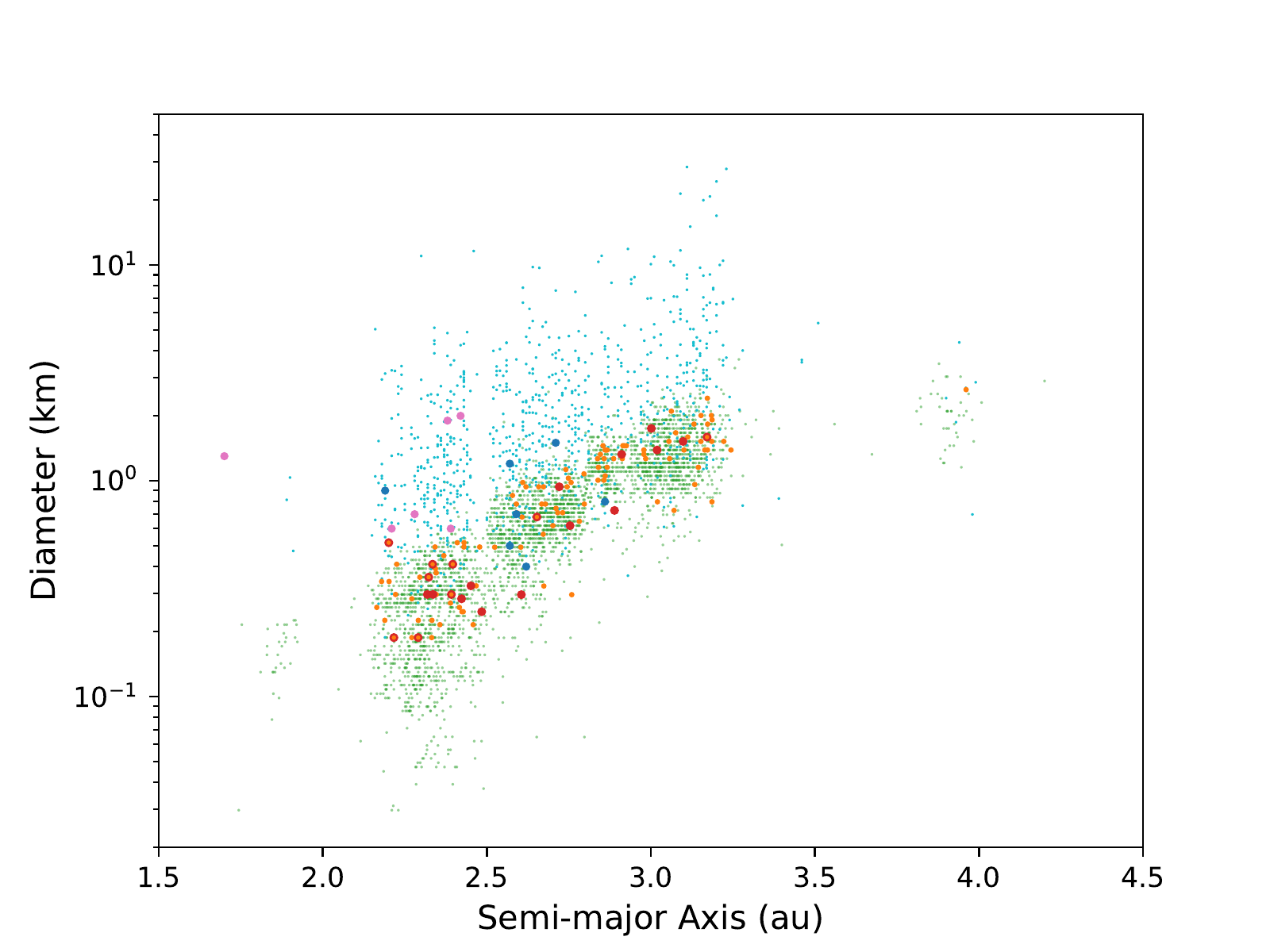}
\caption{Asteroid diameter vs.\ semi-major axis. The light blue dots ad the blue filled circles are the asteroids with reliable rotation periods and the SFRs adopted from \citet{Chang2019}, respectively. The pink filled circles are the other reported SFRs taken from \citet[][see Table 2 therein]{Chang2017}. The green dots are the new asteroids of this work, in which the orange filled circles are the objects with reliable rotation periods and the red filled and open circles are the SFRs and SFR candidates, respectively.}
\label{a_d}
\end{figure}

In total, we found 3574 new asteroids which have five or more detections. Figure~\ref{det_number} shows the distribution of the detection number for these new asteroids, in which the 5 to 10 bin has much more objects. We believe that the high raise in that bin is mainly because some of these short segments were not probably connected in the linkage steps to be identified as the same asteroids, a situation similar the multiple-segment known asteroids mentioned in Section~\ref{sec_ast_rev}. Despite these short-segment asteroids, the high detection number of the new asteroids suggests that they are very unlikely to be false discoveries. Although it is probably difficult to trace them back at this moment due to the orbital uncertainty, we still submitted their detections to the Minor Planet Center in any case of accidental recovery in the near future. The summary table and the entire data set of light curve of these new asteroids can be found in the Table~\ref{table_newast} and~\ref{table_lc}. Figure~\ref{maghist} shows the magnitude distribution of the new asteroids along with the known ones in our observing fields, where the new asteroids are mainly of $21.5 \le w_{p1} \le 22.5$ mag, $\sim 1$ mag fainter than the samples of \citet[][; i.e., roughly 30\% smaller in diameter]{Chang2019}. This suggests that the Hough Transform has pushed the asteroid discovery to the actual limiting magnitude ($\sim 22.5$ mag) of the observation. With this magnitude range, we are able to reach hundred-meter MBAs (see Figure~\ref{a_d}) and have a unique chance to conduct a rotation-period study for them.

In general, the Hough Transform is useful to discover asteroids when applied to the high-cadence observation. Moreover, its implementation is relatively easy and it works effectively on a data set containing massive noises.

\section{Rotation Period Analysis for New PS1 Asteroids}
Out of the 3574 new asteroids, 2853 objects, which have 10 or more detections in $w_{P1}$ band in their light curves, were performed a rotation period analysis using a 2nd-order Fourier series fitting \citep{Harris1989}:
\begin{equation}\label{FTeq}
  M_{i,j} = \sum_{k=1}^{2} B_k\sin\left[\frac{2\pi k}{P} (t_j-t_0)\right] + C_k\cos\left[\frac{2\pi k}{P} (t_j-t_0)\right] + Z_i,
\end{equation}
where $M_{i,j}$ are the reduced magnitudes in $w_{P1}$ band measured at the epoch, $t_j$; $B_k$ and $C_k$ are the coefficients in the Fourier series; $P$ is the rotation period; and $t_0$ is an arbitrary epoch. We also introduced a constant value, $Z_i$, to correct the possible offsets in magnitude between the measurements obtained from different nights.

\begin{figure}
\includegraphics[angle=0,scale=.7]{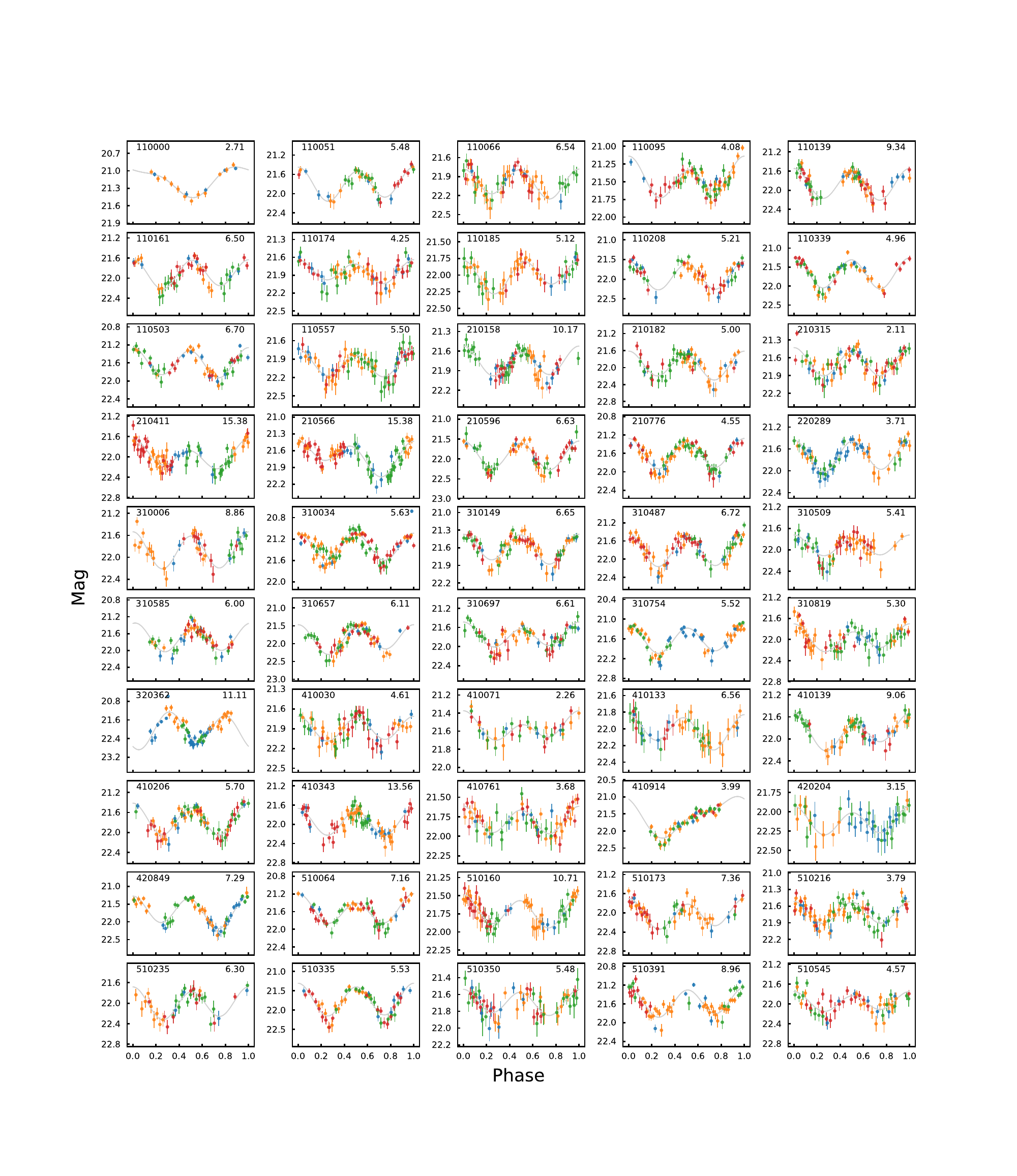}
\caption{Set of 50 folded lightcurves for the reliable rotation periods. Filled circles with different colors are data points in $w_{P1}$ band taken from different nights. The temporary id of the asteroid is given on each plot along with its rotation period in hours.}
\label{lightcurve00}
\end{figure}

\begin{figure}
\includegraphics[angle=0,scale=.7]{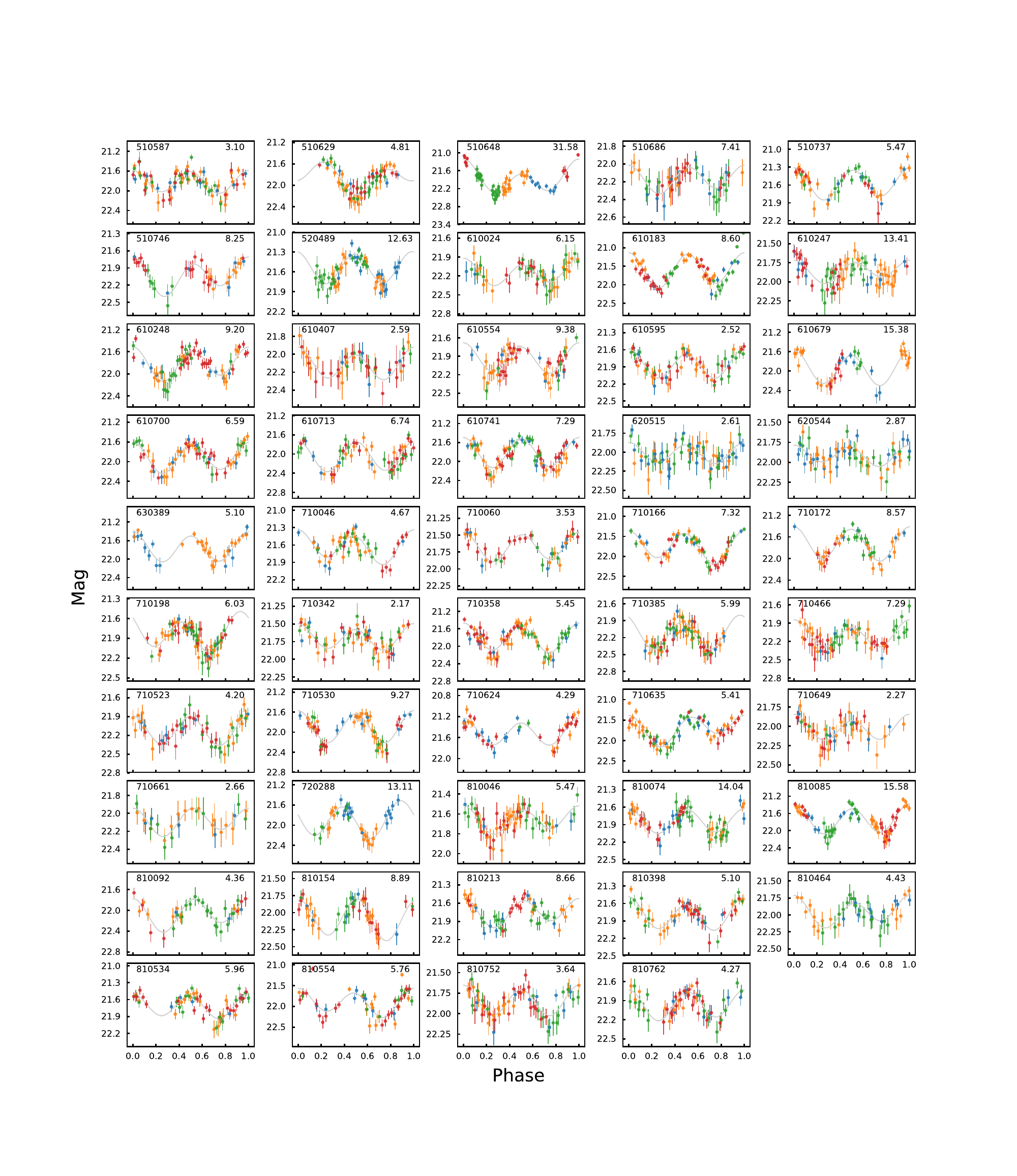}
\caption{Same as Fig.~\ref{lightcurve00} for other 49 reliable rotation periods.}
\label{lightcurve01}
\end{figure}

\begin{figure}
\includegraphics[angle=0,scale=.7]{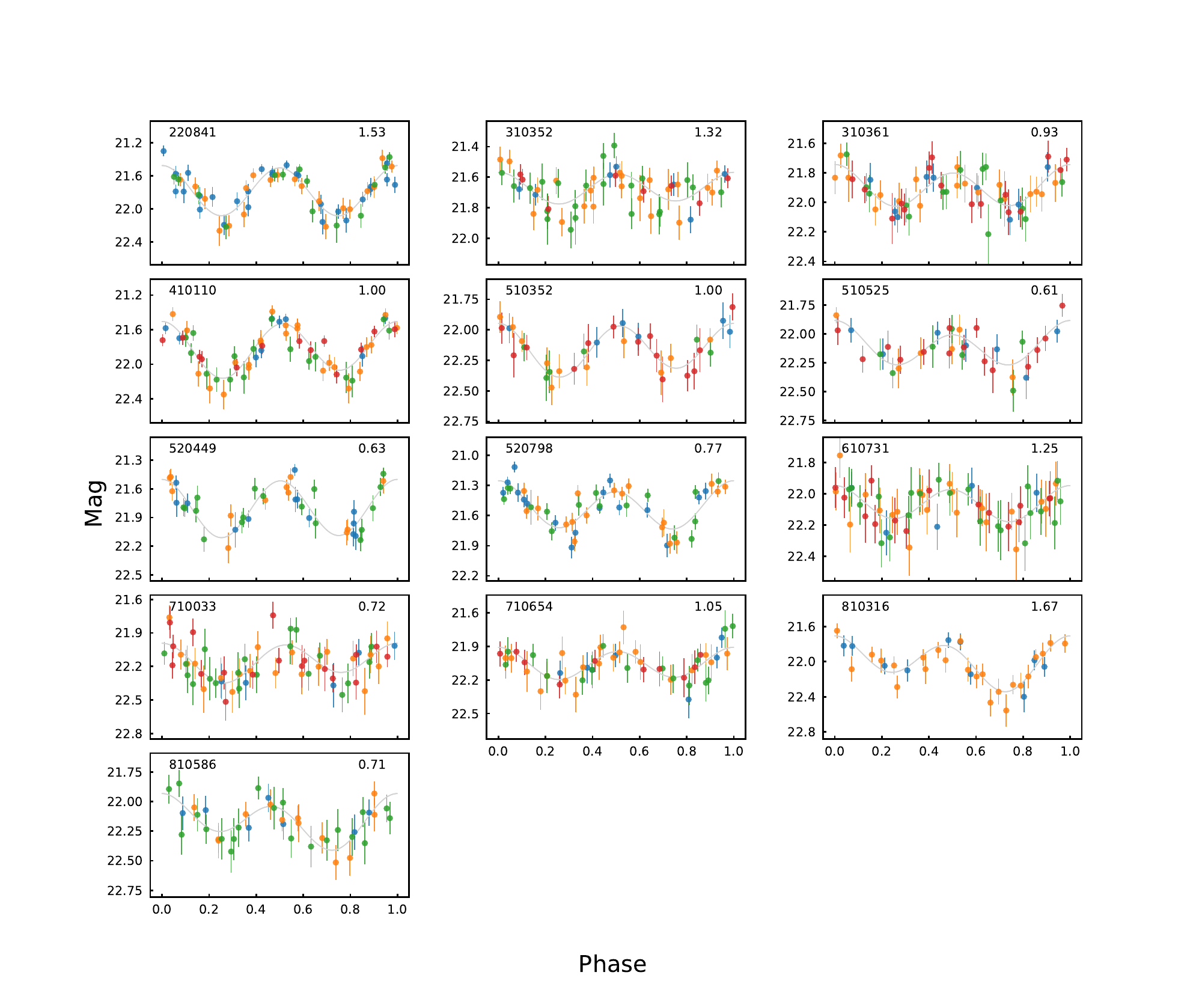}
\caption{The folded light curves of the 13 SFRs. The symbols are the same as Fig.~\ref{lightcurve00}.}
\label{sfr_lc}
\end{figure}

\begin{figure}
\includegraphics[angle=0,scale=.7]{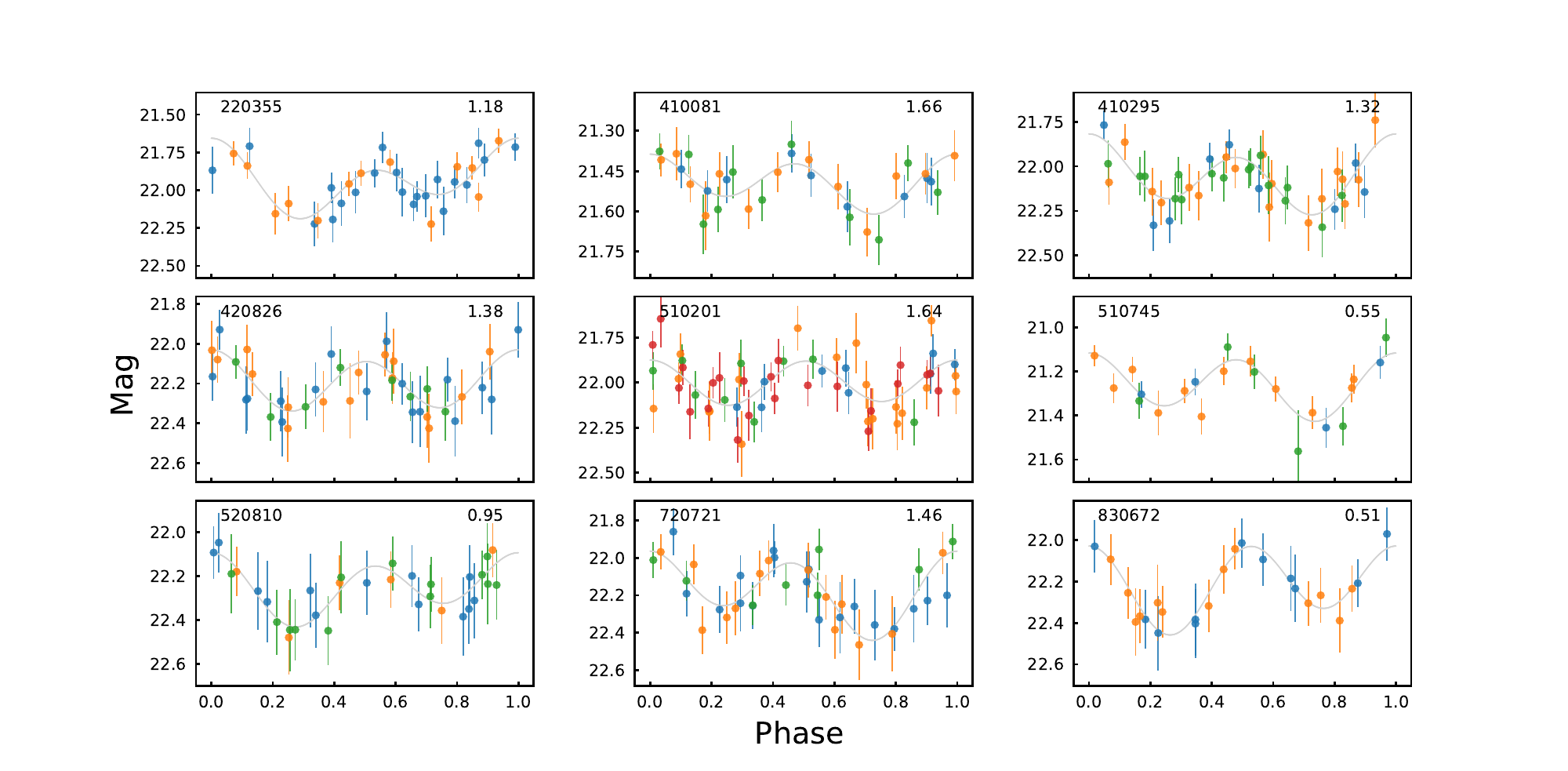}
\caption{The folded light curves of the 9 SFR candidates. The symbols are the same as Fig.~\ref{lightcurve00}.}
\label{sfr_can_lc}
\end{figure}

In the end, we obtained 122 reliable rotation periods. Their information are summarized in the Table~\ref{table_p} and their folded light curves can be found in Figure~\ref{lightcurve00} and~\ref{lightcurve01}. Among the 122 objects, 22 have rotation periods of $< 2$ hr, of which 13 objects were assigned as SFRs due to their highly convincing folded light curves (see Figure~\ref{sfr_lc}) and the other 9 objects were temporarily seen as SFR candidates (see Figure~\ref{sfr_can_lc}). The information of the SFRs and the SFR candidates are listed in Table~\ref{table_sfr} and \ref{table_sfr_can}, respectively.

\begin{figure}
\plotone{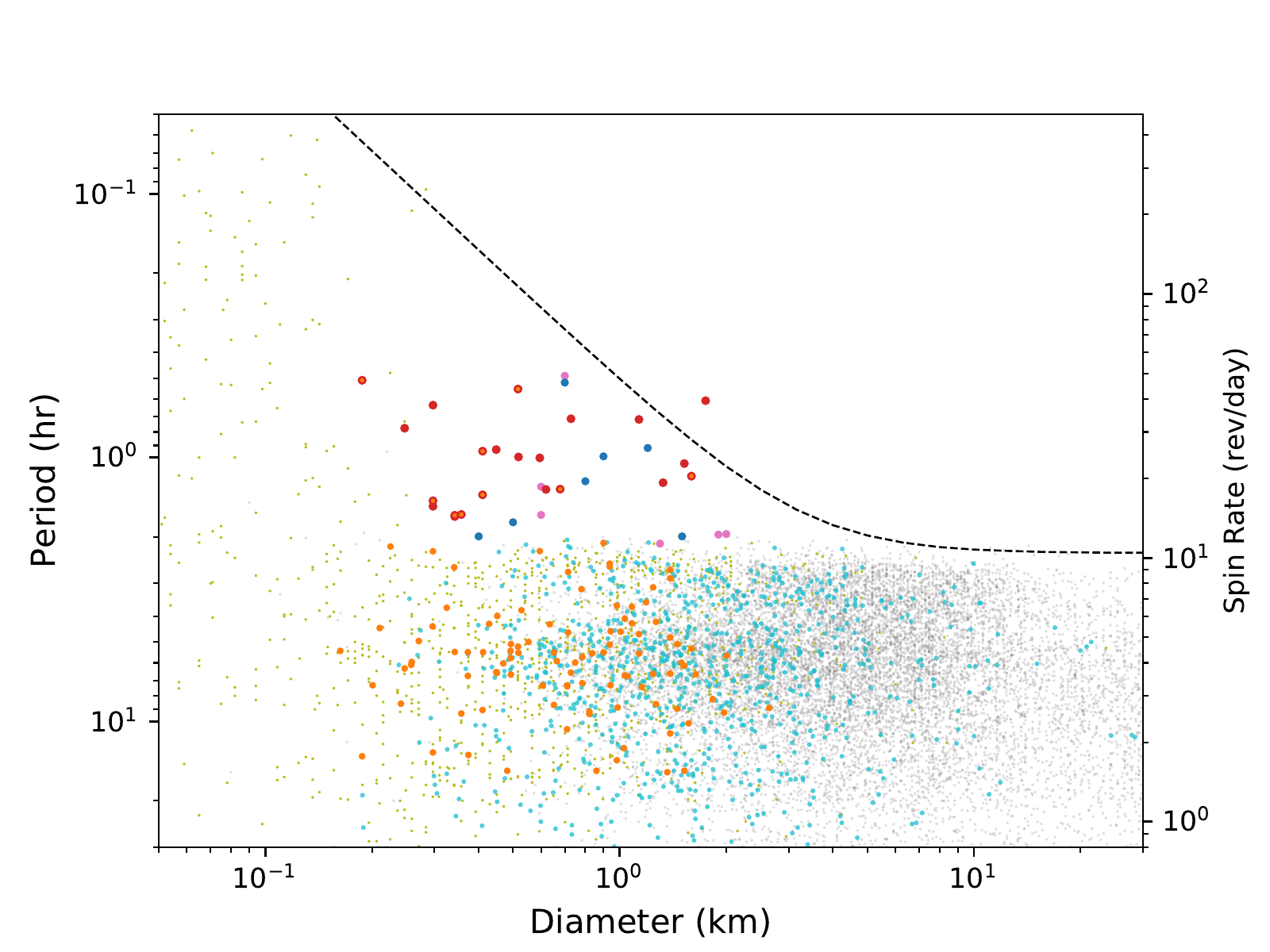}
\caption{Asteroid rotation period vs. diameter. The gray and yellow dots are the asteroids and near-earth asteroids with $U \ge 2$ in the LCDB, respectively. The light blue dots ad the blue filled circles are the asteroids with reliable rotation periods and the SFRs adopted from \citet{Chang2019}, respectively. The pink filled circles are the other reported SFRs taken from \citet[][see Table 2 therein]{Chang2017}. The orange filled circles are the new asteroids with reliable rotation periods of this work, in which the red filled and open circles are the SFRs and SFR candidates, respectively. The dashed line is the spin-rate limit predicted by the size-dependent cohesion model adopted from \citet{Holsapple2007} using $\kappa = 2.25 \times 10^7$ dynes cm$^{-3/2}$.}
\label{dia_per}
\end{figure}

Figure~\ref{dia_per} shows plot of the diameters vs. rotation periods for these 122 objects on top of the samples obtained from \citet{Chang2019} and the $U \ge 2$ objects in the light-curve database \citep[LCDB;][]{Warner2009}. These SFRs are unlikely to be explained by the rubble-pile structure unless they have unusual high bulk density. Compared to the similar studies, e.g., \citet{Chang2015, Chang2016, Chang2019}, the chance of finding SRFs is higher here. As shown in the simulation carried out by \citet[][; see Figure 12 therein]{Chang2019}, the recovery rates are similar for the spin rates of $\ge 3$ rev/day within each magnitude interval\footnote{The simulation of rotation period recovery in \citet{Chang2019} is used to explain our result here because the data sets used in both works are the same.}. This suggests that more SFRs found here is not because we are in favor of detecting short period. Therefore, we believe that small MBAs probably harbor more SFRs. One possible explanation is that these SFRs are monoliths and, moreover, small asteroids are more possible to be monoliths. Consequently, more SFRs could be expected in small MBAs. However, this is difficult to explain why the relatively large SFRs (i.e., diameter of $\sim 1$ km) could avoid numerous collisions after their formations and remain monolithic. Another possible explanation is the size-dependent cohesion model \textbf{in which the cohesion play a significant role for small asteroids, allowing them to rotate faster without breaking apart} \citep{Holsapple2007}. This is another way to expect more SFRs in small asteroids. We adopted the equations shown in \citet{Holsapple2007, Rozitis2014, Polishook2016, Chang2017} and assumed a bulk density of $\rho = 2$ g cm$^{-3}$ to estimate the cohesion required to survive these SFRs under their super-fast spinning. The calculated cohesion of each SFR are given in Table~\ref{table_sfr} and~\ref{table_sfr_can}, in which most cases need a cohesion of tens to thousands of $Pa$, a value consistent with the previously reported SFRs \citep[see Table 4 in ][]{Chang2019} and the regolith on the Moon and Mars \citep[such as the Moon and Mars;][]{Mitchell1974, Sullivan2011}. However, two SFRs, 510525 and 710033, need a cohesion up to thousands of $Pa$, which is several times larger than the average value. This kind of SFRs, larger than kilometer and rotation period of $< 1$ hr, can put a critical constraint on the cohesion model. Therefore, it is encouraged to conduct a comprehensive rotation period survey on the asteroids of few kilometer in size.

\section{Summary}
The intra-night detections of asteroids show up as straight lines and, therefore, the Hough Transform can be used to locate these line-up detections to discover new asteroids. We applied this algorithm to the PS1 high cadence observation, a survey was originally conducted to collect a large sample of asteroid light curves for rotation period measurement during October 26-31, 2016 \citep{Chang2019}. Most of the known asteroids in the observing fields were recovered and, moreover, 3574 new asteroids, mainly of $21.5 < m < 22.5$ mag, were found. Using the light curves of the new asteroids, we obtained 122 reliable rotation periods, of which 13 are SFRs. Compared to the previous surveys, the chance of discovering SFR is much higher here (i.e., $\sim 11\%$). Since the data set used here is the same as \citet{Chang2019}, except our samples are mainly of $21.5 \le w_{p1} \le 22.5$ mag (i.e.,one magnitude fainter than theirs; $\sim 30\%$ smaller in size), our result suggests that sub-kilometer MBAs possibly harbor more SFRs.

\acknowledgments This work is equally contributed by C.-K. Chang and K.-J. Lo. This work is supported by the National Science Council of Taiwan under the grants MOST 107-2112-M-008-009-MY2. We thank the anonymous referee for his comments and suggestions, which makes a great improvement on this work.

\clearpage

\clearpage

\clearpage

\clearpage

\clearpage
\begin{deluxetable}{cccccccc}
\tabletypesize{\scriptsize}
\tablecaption{The PS1 observation in October 2016. \label{obs_log}}
\tablewidth{0pt}
\startdata \tableline\tableline
Field No. & [RA,  Dec.] & Oct 26 2016 & Oct 27 2016 & Oct 28 2014 & Oct 29 2014 & Oct 30 2014 & Oct 31 2014\\
         & [$^{\circ}$,  $^{\circ}$] & $\Delta$t, N$_\textrm{exp}$ & $\Delta$t, N$_\textrm{exp}$ & $\Delta$t, N$_\textrm{exp}$ & $\Delta$t, N$_\textrm{exp}$ & $\Delta$t, N$_\textrm{exp}$ & $\Delta$t, N$_\textrm{exp}$ \\
\tableline
     1 &  [28.10, 13.14] &  6.8, 24 &  6.6, 32 &  6.5, 31  &  5.7, 25 &  6.9, 8 &  1.7, 8 \\
     2 &  [29.17, 10.33] &  6.3, 23 &  6.6, 32 &  6.5, 31  &  5.7, 25 &  6.9, 8 &  1.7, 8 \\
     3 &  [31.00, 14.18] &  6.8, 24 &  6.6, 32 &  6.5, 31  &  5.7, 25 &  6.9, 8 &  1.7, 7 \\
     4 &  [32.04, 11.36] &  6.8, 24 &  6.6, 32 &  6.5, 31  &  5.7, 25 &  6.9, 8 &  1.7, 8 \\
     5 &  [33.92, 15.18] &  6.8, 24 &  6.6, 32 &  6.5, 30  &  5.7, 25 &  6.9, 8 &  1.7, 7 \\
     6 &  [34.93, 12.35] &  6.3, 23 &  6.6, 32 &  6.5, 31  &  5.7, 25 &  6.9, 8 &  1.7, 8 \\
     7 &  [36.87, 16.16] &  6.8, 24 &  6.6, 32 &  6.5, 30  &  5.7, 25 &  6.9, 8 &  1.7, 7 \\
     8 &  [37.85, 13.31] &  6.8, 24 &  6.6, 32 &  6.5, 31  &  5.7, 25 &  6.9, 8 &  1.7, 8 \\
\tableline
\enddata
\tablecomments{$\Delta$t is the time duration spanned by each observing set in hours and
N$_\textrm{exp}$ is the total number of exposures for each night and field. Note that three exposures were taken for each $g_{P1}$, $r_{P1}$, $i_{P1}$, and $z_{P1}$ band on October 26, 2016 and the data sets of last two nights were not used in this work.}
\end{deluxetable}

\begin{deluxetable}{llrrrrrrrrrrlrrrrl}
\tabletypesize{\scriptsize} \setlength{\tabcolsep}{0.02in} \tablecaption{The 122 reliable rotation periods. \label{table_p}} \tablewidth{0pt}
\tablehead{ \colhead{ID} & \colhead{$a$ (au)} & \colhead{$e$} & \colhead{$i$ ($^{\circ}$)} & \colhead{$H_w$ (mag)} & \colhead{D} (km) & \colhead{Period (hr)} & \colhead{$\triangle m$ (mag)} & \colhead{$m_w$ (mag)}}
\startdata
110000 & 2.34$\pm$0.53 & 0.23$\pm$0.20 &  1.08$\pm$ 0.44 &  19.40 &  0.39 &   2.71$\pm$0.02* &  0.48 &  21.19  \\ 
110051 & 2.43$\pm$0.01 & 0.12$\pm$0.01 &  1.84$\pm$ 0.08 &  18.90 &  0.49 &   5.48$\pm$0.04 &  0.70 &  21.74  \\ 
110066 & 3.07$\pm$0.02 & 0.25$\pm$0.01 &  5.85$\pm$ 0.14 &  19.80 &  0.73 &   6.54$\pm$0.07 &  0.57 &  21.97  \\ 
110095 & 2.84$\pm$0.01 & 0.11$\pm$0.01 &  1.07$\pm$ 0.01 &  18.60 &  1.27 &   4.08$\pm$0.02 &  0.51 &  21.49  \\ 
110139 & 2.30$\pm$0.01 & 0.14$\pm$0.01 & 22.79$\pm$ 0.52 &  19.60 &  0.36 &   9.34$\pm$0.07 &  0.67 &  21.81  \\ 
110161 & 2.67$\pm$0.01 & 0.10$\pm$0.01 &  4.43$\pm$ 0.20 &  18.50 &  0.94 &   6.50$\pm$0.05 &  0.72 &  21.89  \\ 
110174 & 2.74$\pm$0.14 & 0.09$\pm$0.13 &  4.98$\pm$ 2.40 &  18.50 &  0.94 &   4.25$\pm$0.03 &  0.61 &  21.86  \\ 
110185 & 2.62$\pm$0.12 & 0.13$\pm$0.12 & 15.83$\pm$ 6.00 &  18.50 &  0.94 &   5.12$\pm$0.04 &  0.54 &  21.96  \\ 
110208 & 2.41$\pm$0.00 & 0.14$\pm$0.02 &  1.04$\pm$ 0.02 &  18.80 &  0.52 &   5.21$\pm$0.02 &  0.84 &  21.83  \\ 
110339 & 2.39$\pm$0.01 & 0.18$\pm$0.01 &  2.50$\pm$ 0.04 &  20.20 &  0.27 &   4.96$\pm$0.02 &  0.95 &  21.65  \\ 
110503 & 2.98$\pm$0.18 & 0.10$\pm$0.14 &  0.59$\pm$ 0.50 &  18.60 &  1.27 &   6.70$\pm$0.04 &  0.77 &  21.62  \\ 
110557 & 2.27$\pm$0.01 & 0.15$\pm$0.01 &  5.28$\pm$ 0.09 &  20.10 &  0.28 &   5.50$\pm$0.05 &  0.59 &  21.98  \\ 
210158 & 3.19$\pm$0.01 & 0.02$\pm$0.01 & 11.33$\pm$ 0.43 &  17.70 &  1.92 &  10.17$\pm$0.18 &  0.50 &  21.80  \\ 
210182 & 2.98$\pm$0.00 & 0.04$\pm$0.02 &  9.52$\pm$ 0.33 &  18.50 &  1.33 &   5.00$\pm$0.03 &  0.71 &  21.91  \\ 
210315 & 2.67$\pm$0.01 & 0.10$\pm$0.01 &  2.24$\pm$ 0.05 &  18.90 &  0.78 &   2.11$\pm$0.01 &  0.61 &  21.70  \\ 
210411 & 3.15$\pm$0.02 & 0.06$\pm$0.03 &  8.20$\pm$ 0.31 &  18.20 &  1.52 &  15.38$\pm$0.30 &  0.71 &  21.97  \\ 
210566 & 3.14$\pm$0.00 & 0.16$\pm$0.01 &  1.29$\pm$ 0.02 &  18.80 &  1.15 &  15.38$\pm$0.30 &  0.64 &  21.68  \\ 
210596 & 3.15$\pm$0.03 & 0.09$\pm$0.02 &  1.25$\pm$ 0.03 &  17.60 &  2.01 &   6.63$\pm$0.06 &  0.84 &  21.91  \\ 
210776 & 2.43$\pm$0.01 & 0.14$\pm$0.01 &  2.10$\pm$ 0.04 &  18.80 &  0.52 &   4.55$\pm$0.02 &  0.74 &  21.64  \\ 
220289 & 2.67$\pm$0.02 & 0.30$\pm$0.00 & 13.96$\pm$ 0.15 &  20.80 &  0.33 &   3.71$\pm$0.01 &  0.66 &  21.72  \\ 
220355 & 3.17$\pm$0.39 & 0.02$\pm$0.16 & 15.97$\pm$ 6.00 &  18.10 &  1.59 &   1.18$\pm$0.00 &  0.49 &  21.95  \\ 
220841 & 2.34$\pm$0.18 & 0.21$\pm$0.12 &  1.14$\pm$ 0.37 &  20.00 &  0.30 &   1.53$\pm$0.00 &  0.75 &  21.78  \\ 
310006 & 2.58$\pm$0.01 & 0.10$\pm$0.02 &  9.50$\pm$ 0.48 &  18.70 &  0.86 &   8.86$\pm$0.13 &  0.66 &  21.81  \\ 
310034 & 3.18$\pm$0.01 & 0.14$\pm$0.01 &  9.32$\pm$ 0.32 &  17.60 &  2.01 &   5.63$\pm$0.03 &  0.67 &  21.34  \\ 
310149 & 2.48$\pm$0.00 & 0.05$\pm$0.01 &  2.68$\pm$ 0.06 &  18.90 &  0.49 &   6.65$\pm$0.04 &  0.61 &  21.58  \\ 
310352 & 2.75$\pm$0.02 & 0.23$\pm$0.01 &  7.14$\pm$ 0.18 &  19.40 &  0.62 &   1.32$\pm$0.00 &  0.38 &  21.69  \\ 
310361 & 2.72$\pm$0.00 & 0.09$\pm$0.02 &  8.85$\pm$ 0.32 &  18.50 &  0.94 &   0.93$\pm$0.00 &  0.41 &  21.92  \\ 
310487 & 2.59$\pm$0.00 & 0.09$\pm$0.01 &  1.32$\pm$ 0.02 &  18.90 &  0.78 &   6.72$\pm$0.04 &  0.80 &  21.72  \\ 
310509 & 2.52$\pm$0.01 & 0.11$\pm$0.01 & 14.09$\pm$ 0.28 &  19.90 &  0.49 &   5.41$\pm$0.04 &  0.60 &  21.94  \\ 
310585 & 3.17$\pm$0.03 & 0.10$\pm$0.02 &  2.16$\pm$ 0.11 &  17.80 &  1.83 &   6.00$\pm$0.03 &  0.75 &  21.70  \\ 
310657 & 2.18$\pm$0.00 & 0.09$\pm$0.01 &  6.26$\pm$ 0.14 &  19.70 &  0.34 &   6.11$\pm$0.05 &  0.88 &  21.84  \\ 
310697 & 3.06$\pm$0.01 & 0.10$\pm$0.01 &  9.12$\pm$ 0.34 &  18.20 &  1.52 &   6.61$\pm$0.07 &  0.66 &  21.82  \\ 
310754 & 5.70$\pm$0.62 & 0.72$\pm$0.04 &  2.10$\pm$ 0.14 &  17.60 &  2.01 &   5.52$\pm$0.03 &  1.04 &  21.57  \\ 
310819 & 3.11$\pm$0.02 & 0.09$\pm$0.02 &  1.39$\pm$ 0.06 &  18.10 &  1.59 &   5.30$\pm$0.04 &  0.69 &  21.97  \\ 
320362 & 3.10$\pm$0.14 & 0.02$\pm$0.10 & 16.36$\pm$ 6.00 &  18.40 &  1.39 &  11.11$\pm$0.16 &  1.35 &  22.01  \\ 
410030 & 2.70$\pm$0.01 & 0.10$\pm$0.01 &  3.07$\pm$ 0.08 &  19.40 &  0.62 &   4.61$\pm$0.03 &  0.54 &  21.91  \\ 
410071 & 2.47$\pm$0.02 & 0.14$\pm$0.02 &  2.43$\pm$ 0.11 &  19.80 &  0.33 &   2.26$\pm$0.01 &  0.38 &  21.58  \\ 
410081 & 2.22$\pm$0.03 & 0.21$\pm$0.01 &  6.50$\pm$ 0.19 &  21.00 &  0.19 &   1.66$\pm$0.01 &  0.26 &  21.49  \\ 
410110 & 2.45$\pm$0.01 & 0.11$\pm$0.01 &  9.94$\pm$ 0.20 &  19.80 &  0.33 &   1.00$\pm$0.00 &  0.72 &  21.82  \\ 
410133 & 2.37$\pm$0.00 & 0.15$\pm$0.02 &  3.60$\pm$ 0.15 &  19.10 &  0.45 &   6.56$\pm$0.09 &  0.43 &  22.01  \\ 
410139 & 2.23$\pm$0.00 & 0.15$\pm$0.01 &  2.46$\pm$ 0.07 &  19.30 &  0.41 &   9.06$\pm$0.10 &  0.66 &  21.83  \\ 
410206 & 3.13$\pm$0.02 & 0.23$\pm$0.01 &  4.13$\pm$ 0.10 &  19.20 &  0.96 &   5.70$\pm$0.03 &  0.72 &  21.76  \\ 
410295 & 2.65$\pm$0.01 & 0.21$\pm$0.01 &  5.73$\pm$ 0.20 &  19.20 &  0.68 &   1.32$\pm$0.00 &  0.45 &  22.09  \\ 
410343 & 2.27$\pm$0.01 & 0.16$\pm$0.00 &  5.35$\pm$ 0.07 &  21.00 &  0.19 &  13.56$\pm$0.23 &  0.75 &  21.93  \\ 
410761 & 2.85$\pm$0.13 & 0.05$\pm$0.13 &  6.21$\pm$ 4.00 &  18.50 &  1.33 &   3.68$\pm$0.02 &  0.46 &  21.80  \\ 
410914 & 2.43$\pm$0.01 & 0.18$\pm$0.00 &  1.15$\pm$ 0.01 &  20.40 &  0.25 &   3.99$\pm$0.04 &  1.06 &  21.70  \\ 
420204 & 2.80$\pm$0.22 & 0.05$\pm$0.14 &  6.66$\pm$ 3.90 &  18.90 &  0.78 &   3.15$\pm$0.03 &  0.47 &  22.12  \\ 
420826 & 2.33$\pm$0.01 & 0.19$\pm$0.03 &  2.11$\pm$ 0.08 &  19.30 &  0.41 &   1.38$\pm$0.01 &  0.41 &  22.21  \\ 
420849 & 2.59$\pm$0.12 & 0.07$\pm$0.10 & 11.87$\pm$ 5.80 &  19.10 &  0.71 &   7.29$\pm$0.04 &  1.02 &  21.73  \\ 
510064 & 2.61$\pm$0.01 & 0.07$\pm$0.01 &  8.09$\pm$ 0.23 &  19.20 &  0.68 &   7.16$\pm$0.06 &  0.82 &  21.61  \\ 
510160 & 3.06$\pm$0.05 & 0.17$\pm$0.06 &  9.26$\pm$ 1.10 &  17.50 &  2.10 &  10.71$\pm$0.19 &  0.50 &  21.73  \\ 
510173 & 2.73$\pm$0.01 & 0.11$\pm$0.01 &  6.66$\pm$ 0.24 &  19.10 &  0.71 &   7.36$\pm$0.09 &  0.75 &  21.99  \\ 
510201 & 2.32$\pm$0.00 & 0.14$\pm$0.01 &  6.62$\pm$ 0.16 &  19.60 &  0.36 &   1.64$\pm$0.01 &  0.45 &  22.02  \\ 
510216 & 2.86$\pm$0.00 & 0.07$\pm$0.01 &  2.13$\pm$ 0.05 &  18.60 &  1.27 &   3.79$\pm$0.02 &  0.65 &  21.71  \\ 
510235 & 2.42$\pm$0.01 & 0.17$\pm$0.01 &  1.83$\pm$ 0.04 &  20.40 &  0.25 &   6.30$\pm$0.07 &  0.69 &  22.01  \\ 
510335 & 2.87$\pm$0.01 & 0.01$\pm$0.02 &  1.39$\pm$ 0.02 &  18.40 &  1.39 &   5.53$\pm$0.01 &  0.91 &  21.83  \\ 
510350 & 2.67$\pm$0.01 & 0.21$\pm$0.01 &  1.45$\pm$ 0.01 &  19.60 &  0.56 &   5.48$\pm$0.05 &  0.43 &  21.71  \\ 
510352 & 2.42$\pm$0.01 & 0.19$\pm$0.01 &  8.38$\pm$ 0.25 &  20.10 &  0.28 &   1.00$\pm$0.00 &  0.47 &  22.15  \\ 
510391 & 2.91$\pm$0.02 & 0.28$\pm$0.01 & 16.19$\pm$ 0.51 &  18.30 &  1.45 &   8.96$\pm$0.14 &  0.77 &  21.59  \\ 
510525 & 3.00$\pm$0.14 & 0.14$\pm$0.14 &  9.27$\pm$ 4.60 &  17.90 &  1.75 &   0.61$\pm$0.00 &  0.43 &  22.13  \\ 
510545 & 2.86$\pm$0.15 & 0.10$\pm$0.13 &  1.47$\pm$ 0.70 &  19.10 &  1.01 &   4.57$\pm$0.04 &  0.60 &  21.99  \\ 
510587 & 3.19$\pm$0.01 & 0.09$\pm$0.01 & 11.67$\pm$ 0.36 &  18.20 &  1.52 &   3.10$\pm$0.01 &  0.69 &  21.80  \\ 
510629 & 2.86$\pm$0.01 & 0.03$\pm$0.01 &  1.83$\pm$ 0.06 &  18.40 &  1.39 &   4.81$\pm$0.04 &  0.68 &  21.92  \\ 
510648 & 3.19$\pm$0.02 & 0.28$\pm$0.00 & 14.67$\pm$ 0.28 &  19.60 &  0.80 &  31.58$\pm$1.75* &  1.21 &  21.95  \\ 
510686 & 2.84$\pm$0.15 & 0.05$\pm$0.13 &  1.35$\pm$ 0.70 &  18.80 &  1.15 &   7.41$\pm$0.16 &  0.41 &  22.20  \\ 
510737 & 2.33$\pm$0.01 & 0.18$\pm$0.00 &  1.13$\pm$ 0.01 &  20.60 &  0.23 &   5.47$\pm$0.03 &  0.61 &  21.54  \\ 
510745 & 2.20$\pm$0.00 & 0.11$\pm$0.03 &  2.52$\pm$ 0.09 &  18.80 &  0.52 &   0.55$\pm$0.00 &  0.37 &  21.27  \\ 
510746 & 3.13$\pm$0.27 & 0.20$\pm$0.16 &  2.31$\pm$ 0.90 &  17.80 &  1.83 &   8.25$\pm$0.11 &  0.54 &  21.98  \\ 
520449 & 2.32$\pm$0.01 & 0.13$\pm$0.01 & 11.40$\pm$ 0.28 &  20.00 &  0.30 &   0.63$\pm$0.00 &  0.65 &  21.79  \\ 
520489 & 2.75$\pm$0.19 & 0.04$\pm$0.13 &  6.47$\pm$ 3.90 &  18.30 &  1.03 &  12.63$\pm$0.27 &  0.55 &  21.61  \\ 
520798 & 2.48$\pm$0.01 & 0.20$\pm$0.00 &  0.57$\pm$ 0.01 &  20.40 &  0.25 &   0.77$\pm$0.00 &  0.62 &  21.51  \\ 
520810 & 2.40$\pm$0.19 & 0.13$\pm$0.15 &  3.11$\pm$ 1.50 &  19.30 &  0.41 &   0.95$\pm$0.01 &  0.35 &  22.27  \\ 
610024 & 3.22$\pm$0.02 & 0.17$\pm$0.02 &  9.44$\pm$ 0.44 &  18.20 &  1.52 &   6.15$\pm$0.13 &  0.57 &  22.15  \\ 
610183 & 3.06$\pm$0.20 & 0.13$\pm$0.15 &  0.80$\pm$ 0.48 &  18.60 &  1.27 &   8.60$\pm$0.09 &  0.98 &  21.67  \\ 
610247 & 2.35$\pm$0.00 & 0.07$\pm$0.01 &  6.48$\pm$ 0.15 &  19.50 &  0.37 &  13.41$\pm$0.47 &  0.47 &  21.88  \\ 
610248 & 2.72$\pm$0.00 & 0.07$\pm$0.01 &  0.75$\pm$ 0.01 &  19.10 &  0.71 &   9.20$\pm$0.11 &  0.65 &  21.82  \\ 
610407 & 2.73$\pm$0.01 & 0.10$\pm$0.02 &  9.67$\pm$ 0.43 &  18.50 &  0.94 &   2.59$\pm$0.01 &  0.40 &  22.11  \\ 
610554 & 2.84$\pm$0.00 & 0.06$\pm$0.02 &  5.94$\pm$ 0.36 &  19.10 &  1.01 &   9.38$\pm$0.15 &  0.58 &  22.02  \\ 
610595 & 2.66$\pm$0.01 & 0.11$\pm$0.01 &  4.46$\pm$ 0.13 &  18.50 &  0.94 &   2.52$\pm$0.01 &  0.58 &  21.85  \\ 
610679 & 2.71$\pm$0.01 & 0.20$\pm$0.01 & 13.79$\pm$ 0.60 &  19.00 &  0.74 &  15.38$\pm$0.40* &  0.74 &  21.86  \\ 
610700 & 3.17$\pm$0.00 & 0.06$\pm$0.02 & 10.14$\pm$ 0.40 &  18.40 &  1.39 &   6.59$\pm$0.05 &  0.73 &  21.87  \\ 
610713 & 2.86$\pm$0.00 & 0.05$\pm$0.01 &  2.97$\pm$ 0.11 &  19.00 &  1.05 &   6.74$\pm$0.06 &  0.74 &  22.03  \\ 
610731 & 2.91$\pm$0.01 & 0.06$\pm$0.01 &  1.58$\pm$ 0.06 &  18.50 &  1.33 &   1.25$\pm$0.01 &  0.40 &  22.08  \\ 
610741 & 2.46$\pm$0.01 & 0.19$\pm$0.00 &  2.31$\pm$ 0.03 &  20.70 &  0.22 &   7.29$\pm$0.04 &  0.76 &  21.87  \\ 
620515 & 2.20$\pm$0.00 & 0.09$\pm$0.01 &  3.64$\pm$ 0.08 &  19.70 &  0.34 &   2.61$\pm$0.01 &  0.48 &  22.04  \\ 
620544 & 2.98$\pm$0.21 & 0.03$\pm$0.15 &  2.06$\pm$ 1.10 &  18.40 &  1.39 &   2.87$\pm$0.02 &  0.39 &  21.91  \\ 
630389 & 2.34$\pm$0.30 & 0.14$\pm$0.16 &  2.78$\pm$ 0.90 &  18.90 &  0.49 &   5.10$\pm$0.07* &  0.68 &  21.74  \\ 
710033 & 3.02$\pm$0.02 & 0.25$\pm$0.01 &  7.59$\pm$ 0.28 &  18.40 &  1.39 &   0.72$\pm$0.00 &  0.56 &  22.18  \\ 
710046 & 2.86$\pm$0.01 & 0.06$\pm$0.02 &  1.93$\pm$ 0.06 &  18.40 &  1.39 &   4.67$\pm$0.04 &  0.62 &  21.63  \\ 
710060 & 2.85$\pm$0.01 & 0.03$\pm$0.01 &  1.77$\pm$ 0.07 &  18.30 &  1.45 &   3.53$\pm$0.02 &  0.48 &  21.67  \\ 
710166 & 2.92$\pm$0.01 & 0.05$\pm$0.01 &  1.53$\pm$ 0.03 &  18.30 &  1.45 &   7.32$\pm$0.04 &  0.85 &  21.75  \\ 
710172 & 2.42$\pm$0.01 & 0.16$\pm$0.00 &  0.69$\pm$ 0.01 &  20.30 &  0.26 &   8.57$\pm$0.12 &  0.67 &  21.78  \\ 
710198 & 3.24$\pm$0.03 & 0.29$\pm$0.01 & 26.82$\pm$ 0.80 &  18.40 &  1.39 &   6.03$\pm$0.03 &  0.55 &  21.90  \\ 
710342 & 2.29$\pm$0.01 & 0.15$\pm$0.00 &  0.80$\pm$ 0.01 &  20.60 &  0.23 &   2.17$\pm$0.01 &  0.49 &  21.70  \\ 
710358 & 2.33$\pm$0.01 & 0.21$\pm$0.00 &  8.94$\pm$ 0.09 &  21.00 &  0.19 &   5.45$\pm$0.04 &  0.83 &  21.81  \\ 
710385 & 2.78$\pm$0.01 & 0.20$\pm$0.01 &  1.90$\pm$ 0.04 &  19.30 &  0.65 &   5.99$\pm$0.05 &  0.64 &  22.21  \\ 
710466 & 2.87$\pm$0.00 & 0.05$\pm$0.01 &  1.02$\pm$ 0.02 &  18.80 &  1.15 &   7.29$\pm$0.09 &  0.48 &  22.14  \\ 
710523 & 2.89$\pm$0.01 & 0.06$\pm$0.01 &  1.64$\pm$ 0.06 &  18.60 &  1.27 &   4.20$\pm$0.03 &  0.54 &  22.10  \\ 
710530 & 3.17$\pm$0.03 & 0.16$\pm$0.04 &  5.19$\pm$ 0.24 &  17.20 &  2.41 &   9.27$\pm$0.11 &  0.69 &  21.94  \\ 
710624 & 2.80$\pm$0.15 & 0.02$\pm$0.14 &  6.13$\pm$ 3.70 &  18.20 &  1.08 &   4.29$\pm$0.02 &  0.69 &  21.45  \\ 
710635 & 2.36$\pm$0.01 & 0.20$\pm$0.00 &  1.49$\pm$ 0.01 &  20.70 &  0.22 &   5.41$\pm$0.02 &  0.87 &  21.73  \\ 
710649 & 2.35$\pm$0.01 & 0.11$\pm$0.01 &  4.39$\pm$ 0.11 &  20.00 &  0.30 &   2.27$\pm$0.01 &  0.47 &  22.04  \\ 
710654 & 3.10$\pm$0.03 & 0.16$\pm$0.01 &  0.46$\pm$ 0.01 &  18.20 &  1.52 &   1.05$\pm$0.00 &  0.43 &  22.06  \\ 
710661 & 3.16$\pm$0.01 & 0.05$\pm$0.03 &  9.97$\pm$ 0.60 &  18.40 &  1.39 &   2.66$\pm$0.04 &  0.40 &  22.09  \\ 
720288 & 2.22$\pm$0.01 & 0.04$\pm$0.01 &  7.76$\pm$ 0.20 &  20.00 &  0.30 &  13.11$\pm$0.22* &  0.72 &  21.92  \\ 
720721 & 2.39$\pm$0.02 & 0.22$\pm$0.02 &  2.84$\pm$ 0.10 &  20.00 &  0.30 &   1.46$\pm$0.01 &  0.43 &  22.17  \\ 
810046 & 2.68$\pm$0.01 & 0.13$\pm$0.01 &  3.60$\pm$ 0.10 &  18.90 &  0.78 &   5.47$\pm$0.05 &  0.36 &  21.67  \\ 
810074 & 2.76$\pm$0.01 & 0.17$\pm$0.01 &  8.84$\pm$ 0.26 &  18.40 &  0.98 &  14.04$\pm$0.25 &  0.62 &  21.83  \\ 
810085 & 3.08$\pm$0.01 & 0.03$\pm$0.01 & 10.43$\pm$ 0.34 &  18.00 &  1.67 &  15.58$\pm$0.21 &  0.84 &  21.77  \\ 
810092 & 2.76$\pm$0.04 & 0.31$\pm$0.00 & 12.51$\pm$ 0.27 &  21.00 &  0.30 &   4.36$\pm$0.04 &  0.64 &  22.04  \\ 
810154 & 3.96$\pm$0.07 & 0.06$\pm$0.04 & 12.89$\pm$ 0.90 &  17.00 &  2.65 &   8.89$\pm$0.17 &  0.53 &  22.00  \\ 
810213 & 3.02$\pm$0.02 & 0.27$\pm$0.01 &  2.48$\pm$ 0.06 &  19.60 &  0.80 &   8.66$\pm$0.13 &  0.57 &  21.75  \\ 
810316 & 2.61$\pm$0.03 & 0.28$\pm$0.02 &  6.32$\pm$ 0.09 &  21.00 &  0.30 &   1.67$\pm$0.01 &  0.63 &  22.03  \\ 
810398 & 2.93$\pm$0.17 & 0.02$\pm$0.15 &  3.13$\pm$ 2.00 &  18.30 &  1.45 &   5.10$\pm$0.03 &  0.55 &  21.75  \\ 
810464 & 2.19$\pm$0.01 & 0.10$\pm$0.01 &  7.91$\pm$ 0.15 &  20.60 &  0.23 &   4.43$\pm$0.05 &  0.53 &  21.96  \\ 
810534 & 2.16$\pm$0.01 & 0.07$\pm$0.00 &  4.39$\pm$ 0.07 &  20.30 &  0.26 &   5.96$\pm$0.04 &  0.52 &  21.67  \\ 
810554 & 2.60$\pm$0.01 & 0.17$\pm$0.01 &  2.53$\pm$ 0.05 &  19.90 &  0.49 &   5.76$\pm$0.06 &  0.84 &  21.93  \\ 
810586 & 2.89$\pm$0.02 & 0.19$\pm$0.01 & 13.45$\pm$ 0.34 &  19.80 &  0.73 &   0.71$\pm$0.00 &  0.53 &  22.18  \\ 
810747 & 2.74$\pm$0.14 & 0.03$\pm$0.14 &  6.25$\pm$ 4.00 &  18.10 &  1.13 &   5.91$\pm$0.04* &  0.89 &  21.38  \\ 
810752 & 2.61$\pm$0.02 & 0.30$\pm$0.01 &  2.88$\pm$ 0.06 &  18.40 &  0.98 &   3.64$\pm$0.02 &  0.44 &  21.88  \\ 
810762 & 2.91$\pm$0.00 & 0.06$\pm$0.01 &  2.11$\pm$ 0.05 &  18.60 &  1.27 &   4.27$\pm$0.02 &  0.56 &  21.94  \\ 
830672 & 2.29$\pm$0.31 & 0.17$\pm$0.17 &  3.60$\pm$ 2.70 &  21.00 &  0.19 &   0.51$\pm$0.00 &  0.39 &  22.25  \\ 
\enddata
\tablecomments{Columns: temporary id, semi-major axis ($a$, AU), eccentricity ($e$,
degree), inclination ($i$, degree), absolute magnitude in $w_{p1}$ band ($H$, mag), diameter ($D$, km),
derived rotation period (hours), lightcurve amplitude (mag), and mean apparent magnitude in $w_{p1}$ band (mag).
Note that half-periods are marked with ``*''.}
\end{deluxetable}

\begin{deluxetable}{llrrrrrrrrrrlrrrrlr}
\tabletypesize{\scriptsize} \setlength{\tabcolsep}{0.02in} \tablecaption{The 13 super-fast rotators.. \label{table_sfr}} \tablewidth{0pt}
\tablehead{ \colhead{ID} & \colhead{$a$ (au)} & \colhead{$e$} & \colhead{$i$ ($^{\circ}$)} & \colhead{$H_w$ (mag)} & \colhead{D} (km) & \colhead{Period (hr)} & \colhead{$\triangle m$ (mag)} & \colhead{$m_w$ (mag)}}
\startdata
220841 & 2.34$\pm$0.18 & 0.21$\pm$0.12 &  1.14$\pm$ 0.37 &  20.00 &  0.30 &   1.53$\pm$0.00 &  0.75 &  21.78 &    20 \\ 
310352 & 2.75$\pm$0.02 & 0.23$\pm$0.01 &  7.14$\pm$ 0.18 &  19.40 &  0.62 &   1.32$\pm$0.00 &  0.38 &  21.69 &    80 \\ 
310361 & 2.72$\pm$0.00 & 0.09$\pm$0.02 &  8.85$\pm$ 0.32 &  18.50 &  0.94 &   0.93$\pm$0.00 &  0.41 &  21.92 &   438 \\ 
410110 & 2.45$\pm$0.01 & 0.11$\pm$0.01 &  9.94$\pm$ 0.20 &  19.80 &  0.33 &   1.00$\pm$0.00 &  0.72 &  21.82 &    64 \\ 
510352 & 2.42$\pm$0.01 & 0.19$\pm$0.01 &  8.38$\pm$ 0.25 &  20.10 &  0.28 &   1.00$\pm$0.00 &  0.47 &  22.15 &    36 \\ 
510525 & 3.00$\pm$0.14 & 0.14$\pm$0.14 &  9.27$\pm$ 4.60 &  17.90 &  1.75 &   0.61$\pm$0.00 &  0.43 &  22.13 &  3880 \\ 
520449 & 2.32$\pm$0.01 & 0.13$\pm$0.01 & 11.40$\pm$ 0.28 &  20.00 &  0.30 &   0.63$\pm$0.00 &  0.65 &  21.79 &   132 \\ 
520798 & 2.48$\pm$0.01 & 0.20$\pm$0.00 &  0.57$\pm$ 0.01 &  20.40 &  0.25 &   0.77$\pm$0.00 &  0.62 &  21.51 &    57 \\ 
610731 & 2.91$\pm$0.01 & 0.06$\pm$0.01 &  1.58$\pm$ 0.06 &  18.50 &  1.33 &   1.25$\pm$0.01 &  0.40 &  22.08 &   438 \\ 
710033 & 3.02$\pm$0.02 & 0.25$\pm$0.01 &  7.59$\pm$ 0.28 &  18.40 &  1.39 &   0.72$\pm$0.00 &  0.56 &  22.18 &  1995 \\ 
710654 & 3.10$\pm$0.03 & 0.16$\pm$0.01 &  0.46$\pm$ 0.01 &  18.20 &  1.52 &   1.05$\pm$0.00 &  0.43 &  22.06 &   889 \\ 
810316 & 2.61$\pm$0.03 & 0.28$\pm$0.02 &  6.32$\pm$ 0.09 &  21.00 &  0.30 &   1.67$\pm$0.01 &  0.63 &  22.03 &    13 \\ 
810586 & 2.89$\pm$0.02 & 0.19$\pm$0.01 & 13.45$\pm$ 0.34 &  19.80 &  0.73 &   0.71$\pm$0.00 &  0.53 &  22.18 &   537 \\ 
\enddata
\tablecomments{Columns: temporary id, semi-major axis ($a$, AU), eccentricity ($e$,
degree), inclination ($i$, degree), absolute magnitude in $w_{p1}$ band ($H$, mag), diameter ($D$, km),
derived rotation period (hours), lightcurve amplitude (mag), mean apparent magnitude in $w_{p1}$ band (mag), and estimated cohesion ($Pa$).}
\end{deluxetable}

\begin{deluxetable}{llrrrrrrrrrrlrrrrl}
\tabletypesize{\scriptsize} \setlength{\tabcolsep}{0.02in} \tablecaption{The 9 super-fast-rotator candidates. \label{table_sfr_can}} \tablewidth{0pt}
\tablehead{ \colhead{ID} & \colhead{$a$ (au)} & \colhead{$e$} & \colhead{$i$ ($^{\circ}$)} & \colhead{$H_w$} & \colhead{D} (km) & \colhead{Period (hr)} & \colhead{$\triangle m$ (mag)} & \colhead{$m_w$ (mag)}}
\startdata
220355 & 3.17$\pm$0.39 & 0.02$\pm$0.16 & 15.97$\pm$ 6.00 &  18.10 &  1.59 &   1.18$\pm$0.00 &  0.49 &  21.95   &   808 \\ 
410081 & 2.22$\pm$0.03 & 0.21$\pm$0.01 &  6.50$\pm$ 0.19 &  21.00 &  0.19 &   1.66$\pm$0.01 &  0.26 &  21.49   &     3 \\ 
410295 & 2.65$\pm$0.01 & 0.21$\pm$0.01 &  5.73$\pm$ 0.20 &  19.20 &  0.68 &   1.32$\pm$0.00 &  0.45 &  22.09   &   106 \\ 
420826 & 2.33$\pm$0.01 & 0.19$\pm$0.03 &  2.11$\pm$ 0.08 &  19.30 &  0.41 &   1.38$\pm$0.01 &  0.41 &  22.21   &    32 \\ 
510201 & 2.32$\pm$0.00 & 0.14$\pm$0.01 &  6.62$\pm$ 0.16 &  19.60 &  0.36 &   1.64$\pm$0.01 &  0.45 &  22.02   &    15 \\ 
510745 & 2.20$\pm$0.00 & 0.11$\pm$0.03 &  2.52$\pm$ 0.09 &  18.80 &  0.52 &   0.55$\pm$0.00 &  0.37 &  21.27   &   394 \\ 
520810 & 2.40$\pm$0.19 & 0.13$\pm$0.15 &  3.11$\pm$ 1.50 &  19.30 &  0.41 &   0.95$\pm$0.01 &  0.35 &  22.27   &    76 \\ 
720721 & 2.39$\pm$0.02 & 0.22$\pm$0.02 &  2.84$\pm$ 0.10 &  20.00 &  0.30 &   1.46$\pm$0.01 &  0.43 &  22.17   &    15 \\ 
830672 & 2.29$\pm$0.31 & 0.17$\pm$0.17 &  3.60$\pm$ 2.70 &  21.00 &  0.19 &   0.51$\pm$0.00 &  0.39 &  22.25   &    62 \\ 
\enddata
\tablecomments{Columns are the same with Table~\ref{table_sfr}.}
\end{deluxetable}

\begin{deluxetable}{lcccrr}
\tabletypesize{\scriptsize} \setlength{\tabcolsep}{0.02in} \tablecaption{The information of new asteroids. \label{table_newast}} \tablewidth{0pt}
\tablehead{ \colhead{ID} & \colhead{$a$ (au)} & \colhead{$e$} & \colhead{$i$ ($^\circ$)} & \colhead{$H_w$ (mag)} & \colhead{D (km)}}
\startdata
110000 & 2.34$\pm$0.53 & 0.23$\pm$0.20 &  1.08$\pm$ 0.44 &  19.40 &  0.39 \\
110006 & 2.33$\pm$0.01 & 0.18$\pm$0.00 &  1.92$\pm$ 0.03 &  20.90 &  0.20 \\
110007 & 3.11$\pm$0.01 & 0.12$\pm$0.01 &  9.73$\pm$ 0.34 &  18.60 &  1.27 \\
110012 & 2.78$\pm$0.19 & 0.10$\pm$0.11 &  6.90$\pm$ 4.30 &  19.10 &  0.71 \\
110015 & 2.54$\pm$0.00 & 0.04$\pm$0.01 &  3.53$\pm$ 0.10 &  19.60 &  0.56 \\
110020 & 2.58$\pm$0.02 & 0.26$\pm$0.00 &  4.72$\pm$ 0.09 &  21.10 &  0.28 \\
110022 & 2.66$\pm$0.01 & 0.09$\pm$0.01 &  2.60$\pm$ 0.08 &  19.70 &  0.54 \\
110023 & 2.32$\pm$0.89 & 0.39$\pm$0.25 &  5.62$\pm$38.00 &  20.90 &  0.20 \\
110025 & 2.39$\pm$0.02 & 0.19$\pm$0.02 &  0.96$\pm$ 0.04 &  19.40 &  0.39 \\
110029 & 2.55$\pm$0.01 & 0.20$\pm$0.02 &  3.90$\pm$ 0.13 &  19.00 &  0.74 \\
110037 & 3.13$\pm$0.02 & 0.02$\pm$0.02 &  3.80$\pm$ 0.18 &  17.80 &  1.83 \\
110038 & 2.37$\pm$0.00 & 0.19$\pm$0.02 &  5.76$\pm$ 0.24 &  18.70 &  0.54 \\
110045 & 2.51$\pm$0.61 & 0.03$\pm$0.21 &  4.05$\pm$ 3.50 &  19.60 &  0.56 \\
110046 & 2.76$\pm$0.14 & 0.06$\pm$0.13 &  2.34$\pm$ 1.50 &  19.00 &  0.74 \\
110051 & 2.43$\pm$0.01 & 0.12$\pm$0.01 &  1.84$\pm$ 0.08 &  18.90 &  0.49 \\
110063 & 3.04$\pm$0.19 & 0.11$\pm$0.13 &  1.49$\pm$ 1.20 &  19.20 &  0.96 \\
110066 & 3.07$\pm$0.02 & 0.25$\pm$0.01 &  5.85$\pm$ 0.14 &  19.80 &  0.73 \\
110069 & 3.14$\pm$0.01 & 0.16$\pm$0.03 & 10.22$\pm$ 0.70 &  18.00 &  1.67 \\
110071 & 2.54$\pm$0.00 & 0.05$\pm$0.01 &  0.77$\pm$ 0.01 &  19.50 &  0.59 \\
110077 & 2.88$\pm$1.15 & 0.05$\pm$0.27 &  1.41$\pm$70.00 &  18.40 &  1.39 \\
\enddata
\tablecomments{Columns: temporary id, semi-major axis ($a$, au), eccentricity ($e$), inclination ($i$, degree), absolute magnitude in $w_{P1}$ band (mag), and calculated diameter (km). Note that the full table is available for the online version.}
\end{deluxetable}

\begin{deluxetable}{llllcc}
\tabletypesize{\scriptsize} \setlength{\tabcolsep}{0.02in} \tablecaption{The light-curve data set. \label{table_lc}} \tablewidth{0pt}
\tablehead{ \colhead{ID} & \colhead{MJD} & \colhead{R.A. (hms)} & \colhead{Dec. (dms)} & \colhead{$m$ (mag)} & \colhead{Filter}}
\startdata
 110000 & 57687.29448 & 01 50 20.990 & +12 38 28.43 & 21.92 $\pm$ 0.12 & g \\
 110000 & 57687.30355 & 01 50 20.415 & +12 38 25.07 & 21.25 $\pm$ 0.05 & w \\
 110000 & 57687.31329 & 01 50 19.806 & +12 38 21.58 & 21.51 $\pm$ 0.08 & r \\
 110000 & 57687.32231 & 01 50 19.221 & +12 38 18.20 & 21.18 $\pm$ 0.05 & w \\
 110000 & 57687.33210 & 01 50 18.598 & +12 38 14.66 & 20.92 $\pm$ 0.06 & i \\
 110000 & 57687.34120 & 01 50 18.015 & +12 38 11.27 & 20.83 $\pm$ 0.04 & w \\
 110000 & 57687.35409 & 01 50 17.167 & +12 38 06.42 & 20.63 $\pm$ 0.05 & z \\
 110000 & 57687.37634 & 01 50 15.737 & +12 37 58.33 & 21.28 $\pm$ 0.06 & g \\
 110000 & 57687.38548 & 01 50 15.179 & +12 37 55.01 & 20.91 $\pm$ 0.04 & w \\
 110000 & 57687.39549 & 01 50 14.536 & +12 37 51.30 & 21.07 $\pm$ 0.05 & r \\
 110000 & 57687.46472 & 01 50 10.089 & +12 37 25.56 & 20.80 $\pm$ 0.04 & w \\
 110000 & 57687.47452 & 01 50 09.459 & +12 37 21.86 & 20.67 $\pm$ 0.05 & i \\
 110000 & 57688.28726 & 01 49 20.322 & +12 32 20.68 & 21.03 $\pm$ 0.05 & w \\
 110000 & 57688.29380 & 01 49 19.916 & +12 32 18.27 & 21.15 $\pm$ 0.06 & w \\
 110000 & 57688.30033 & 01 49 19.503 & +12 32 15.93 & 21.14 $\pm$ 0.05 & w \\
 110000 & 57688.30687 & 01 49 19.090 & +12 32 13.63 & 21.24 $\pm$ 0.05 & w \\
 110000 & 57688.31342 & 01 49 18.682 & +12 32 11.20 & 21.33 $\pm$ 0.06 & w \\
 110000 & 57688.31996 & 01 49 18.256 & +12 32 08.73 & 21.46 $\pm$ 0.07 & w \\
 110000 & 57688.32652 & 01 49 17.859 & +12 32 06.31 & 21.53 $\pm$ 0.08 & w \\
 110000 & 57688.33311 & 01 49 17.414 & +12 32 04.14 & 21.42 $\pm$ 0.07 & w \\
 110000 & 57688.33998 & 01 49 16.996 & +12 32 01.26 & 21.40 $\pm$ 0.07 & w \\
 110000 & 57688.35428 & 01 49 16.085 & +12 31 56.36 & 21.07 $\pm$ 0.04 & w \\
 110000 & 57688.36090 & 01 49 15.672 & +12 31 54.00 & 21.00 $\pm$ 0.05 & w \\
 110000 & 57688.36753 & 01 49 15.246 & +12 31 51.41 & 20.91 $\pm$ 0.05 & w \\
\enddata
\tablecomments{Columns: temporary id, observation epoch (mjd), right ascension (hms), declination (dms), magnitude, and PS1 filter. Note that the full table is available for the online version.}
\end{deluxetable}

\end{document}